\RequirePackage{fix-cm}

\documentclass[twocolumn]{svjour3}          
\smartqed  
\usepackage{graphicx}
\usepackage{caption}
\usepackage{amsmath}
\usepackage{amsfonts}
\usepackage{subcaption}
\usepackage{hyperref}
\usepackage{color}
\usepackage{stfloats}
\usepackage{floatrow}
\usepackage{multirow}
\usepackage{ctable}
\usepackage{soul}
\journalname{Nonlinear Dynamics}

\begin{document}

\title{Global sensitivity analysis of asymmetric energy harvesters}

\author{Jo\~ao Pedro Norenberg         \and
        Americo Cunha Jr               \and
        Samuel da Silva                \and
        \\Paulo Sergio Varoto
}


\institute{Jo\~ao Pedro Norenberg \at
           S\~ao Paulo State University, Ilha Solteira, SP, Brazil  \\
           ORCID: 0000-0003-3558-4053\\
           \email{jp.norenberg@unesp.br}
           \and
           Americo Cunha Jr \at
           Rio de Janeiro State University, Rio de Janeiro, RJ, Brazil\\
           ORCID: 0000-0002-8342-0363\\
           \email{americo.cunha@uerj.br}
           \and
           Samuel da Silva \at
           S\~ao Paulo State University, Ilha Solteira, SP, Brazil  \\
           ORCID: 0000-0001-6430-3746\\
           \email{samuel.silva13@unesp.br}
           \and
           Paulo Sergio Varoto \at
           University of S\~ao Paulo, S\~ao Carlos, SP, Brazil \\
           ORCID: 0000-0002-1240-1720\\
           \email{varoto@sc.usp.br}
}

\date{Received: date / Accepted: date}

\maketitle
\begin{abstract}
Parametric variability is inevitable in actual energy harvesters. It can significantly affect crucial aspects of the system performance, especially in harvesting systems that present geometric parameters,  material properties, or excitation conditions that are susceptible to small perturbations. This work aims to develop an investigation to identify the most critical parameters in the dynamic behavior of asymmetric bistable energy harvesters with nonlinear piezoelectric coupling, considering the variability of their physical and excitation properties. For this purpose, a global sensitivity analysis based on orthogonal variance decomposition, employing Sobol indices, is performed to quantify the effect of the harvester parameters on the variance of the recovered power. This technique quantifies the variance concerning each parameter individually and collectively regarding the total variation of the model. The results indicate that the frequency and amplitude of excitation, asymmetric terms and electrical proprieties of the piezoelectric coupling are the most critical parameters that affect the mean power harvested. It is also shown that the order of importance of the parameters can change according to the stability of the harvester's dynamic response. In this way, a better understanding of the system under analysis is obtained since the study allows the identification of vital parameters that rule the change of dynamic behavior and therefore constitutes a powerful tool in the robust design, optimization, and response prediction of nonlinear harvesters.

\end{abstract}

\section{Introduction}
\label{intro}

Novel technological applications that use small-scale autonomous devices continue to emerge every year. In this context, one of the most significant challenges in developing a self-sufficient machine, i.e., an electromechanical system capable of generating electrical power to sustain its operation or provide additional energy supply. In this perspective, energy harvesting has become a promising research area that comprises dispersed energy in the environment (e.g., solar, wind, heat, vibration, etc.) to convert into electricity using a suitable conversion process. 

Among commonly available energy sources present in the environment, the kinetic energy from environmental noise and structural vibration signals, usually wasted, has been major explored for energy harvesting due to a good compromise between energy availability and ease of use. The literature has several recent works with powerful applications involving this idea, such as the generation of bio-cellular energy \cite{1_catacuzzeno}, medical implants \cite{2_inman}, sensor power \cite{4_Lee,3_wang,5_fang}, and even osmotic energy \cite{6_Xin}. The conversion techniques used in vibration energy harvesting are electrostatic \cite{20_Mitcheson}, electromagnetism \cite{21_Arnold}, and piezoelectricity \cite{22_Erturk}. Piezoelectricity, in particular, has received more attention because of its simplicity and the high-density energy of the resulting electrical signals when the piezoelectric material is used in the sensing mode of operation, which is typically the case in piezoelectric energy harvesting systems.

In a broad sense, energy harvesting devices are modeled as deterministic systems, and, consequently, all corresponding model parameters are free of uncertainty. However, even the most accurate manufacturing processes can present some degree of geometrical variability in large-scale production. Additionally, material property variability can also occur. These, along with other environmental factors, are sources of uncertainty and can affect the overall performance of a given energy harvesting system. Thus, disregarding these sources of uncertainties represents a significant limitation of the deterministic approach, leading to low fidelity or even wrong predictions. Nowadays, all engineering fields that deal with robust and accurate predictions require suitable methods to characterize and quantify these uncertainties, and the general area from applied mathematics and statistics that studies the effect of uncertainties is usually known as Uncertainty Quantification (UQ) \cite{Soize2017}. This relevant research relies on stochastic techniques to determine the statistical output distribution given the corresponding system's input distribution. UQ is essential in energy harvesting since it can handle low-scale and accurate models. Recent reviews of UQ in the context of energy harvesting are found in \cite{18_Franco,9_Li,10_Mann}.

Sensitivity analysis (SA) is a tool that has been recently receiving increasing attention in the analysis of many engineering systems since it offers a comprehensive evaluation of how the system properties can affect the system's response \cite{27_Abbiati}. It allows for identifying the most critical parameters that affect the energy harvesting dynamic response. According to \cite{24_Saltelli}, SA aims to determine which factors require a more precise estimation and to identify the weak links in the evaluation chain (those that propagate the most significant variance in the output). Besides assisting UQ techniques in formulating a more straightforward probabilistic model for the system of interest, this analysis can also bring further insight into the system's behavior.

SA techniques can be classified as local and global \cite{23_Cacuci}. The first is based on the approximation of the partial derivative to assess how the variation in one parameter affects the system's response while keeping other parameters fixed according to their nominal values. The second approach provides a comprehensive analysis, evaluating the effect of simultaneously varying the parameters within the multidimensional input space. A helpful survey on the state-of-the-art sensitivity analysis techniques is available at \cite{24_Saltelli}.

Applications of SA on energy harvesting can be seen, for instance, in Aloui, Larbi, and Chouchane \cite{26_Aloui,Aloui_2020}, where the authors applied a global sensitivity analysis (GSA) on a linear bimorph piezoelectric energy harvester subjected to a base excitation with a mounted load resistance in series. Ruiz and Meruane \cite{28_Ruiz} conducted GSA based on a variance method concerning the behavior of the frequency response function of the linear energy harvester under unimorph and bimorph piezoelectric configurations. These studies concluded that the effects of physical properties of the piezoelectric materials are fundamental for the understanding of variabilities in the models, and they draw attention to the need to investigate the propagation of uncertainties in piezoelectric energy harvesters. However, to the author's knowledge, there is a gap in the literature for sensitivity analysis in nonlinear energy harvesting systems that could provide in quantitative terms which parameters most affect the variability of the system's response.

Linear models have limited bandwidth around their fundamental resonance frequency. This limitation can be overcome by exploiting nonlinearities, which allow energy scavenges in broadband frequency. Cottone et al. \cite{7_Cottone} and Erturk et al. \cite{8_Erturk} were the first works to implement nonlinear bistable energy harvesting systems. They demonstrated a substantial energy increase compared to the linear system in a broadband scenario. This energy harvester has become a classic system, being studied by several authors \cite{16_Cao,CunhaJR_CE,Daqaq,19_dAQAQ,Erturk_valid,17_Huang,Lopes,Yang_Fei} due to its rich dynamics and complexity. In addition to these studies, nonlinear energy harvesting systems can be exploited via SA, which can be seen as an exploratory tool capable of revealing individual and joint parametric contributions to the system response. Since nonlinear systems present high sensitivity to input parameter variations and may have complex dynamics with regular and chaotic behaviors and multiple solutions, SA may reveal nontrivial input-output relationships that can improve the energy recovery process.

This work is concerned with the global sensitivity analysis of bistable energy harvesting systems. Several sources of uncertainties and nonlinearities are considered to explore the physical system behavior, such as multiple excitation conditions and geometric asymmetries. Those asymmetries can arise from the manufacturing process, the buckling eccentricity, magnetic forces, and heterogeneity of the materials used. Therefore, accounting for these effects leads to more reliable models for assessing the system's performance in the electromechanical conversion process. Hence, a probabilistic model of uncertainties employing a suitable joint probability distribution for the model parameters is conceived. The goal here is to classify which selected parameters most affect the harvested power. The model allows a clear understanding of the physical phenomena involved in the bistable energy harvesting system dynamics. Mainly through the interactions between the corresponding variables, which is an essential step towards reaching robustness of the model employed, serving as a training tool to determine the essential features closely linked to the harvesting output response data.

The remaining parts of this manuscript are organized as follows. 
Section~\ref{sec:bistable energy harvesting} presents the energy harvesting system of interest. The global sensitivity analysis of this nonlinear system is formulated in Section~\ref{sec:Global sensitivity}, and the Polynomial Chaos Expansion approach used to compute the Sobol indices is presented in Section~\ref{sec:PCE}. In Section~\ref{sec:NE}, the numerical experiments conducted to analyze the sensitivity of a bistable energy harvester are presented and discussed. Section~\ref{sec:summary} summarizes the main results obtained here and the uncertainties propagation of the systems. Finally, in Section~\ref{sec:final remarks}, important conclusions are presented.

\section{Bistable energy harvesting system}
\label{sec:bistable energy harvesting}

\begin{figure}[hbt]
    \centering
    \includegraphics[scale=0.4]{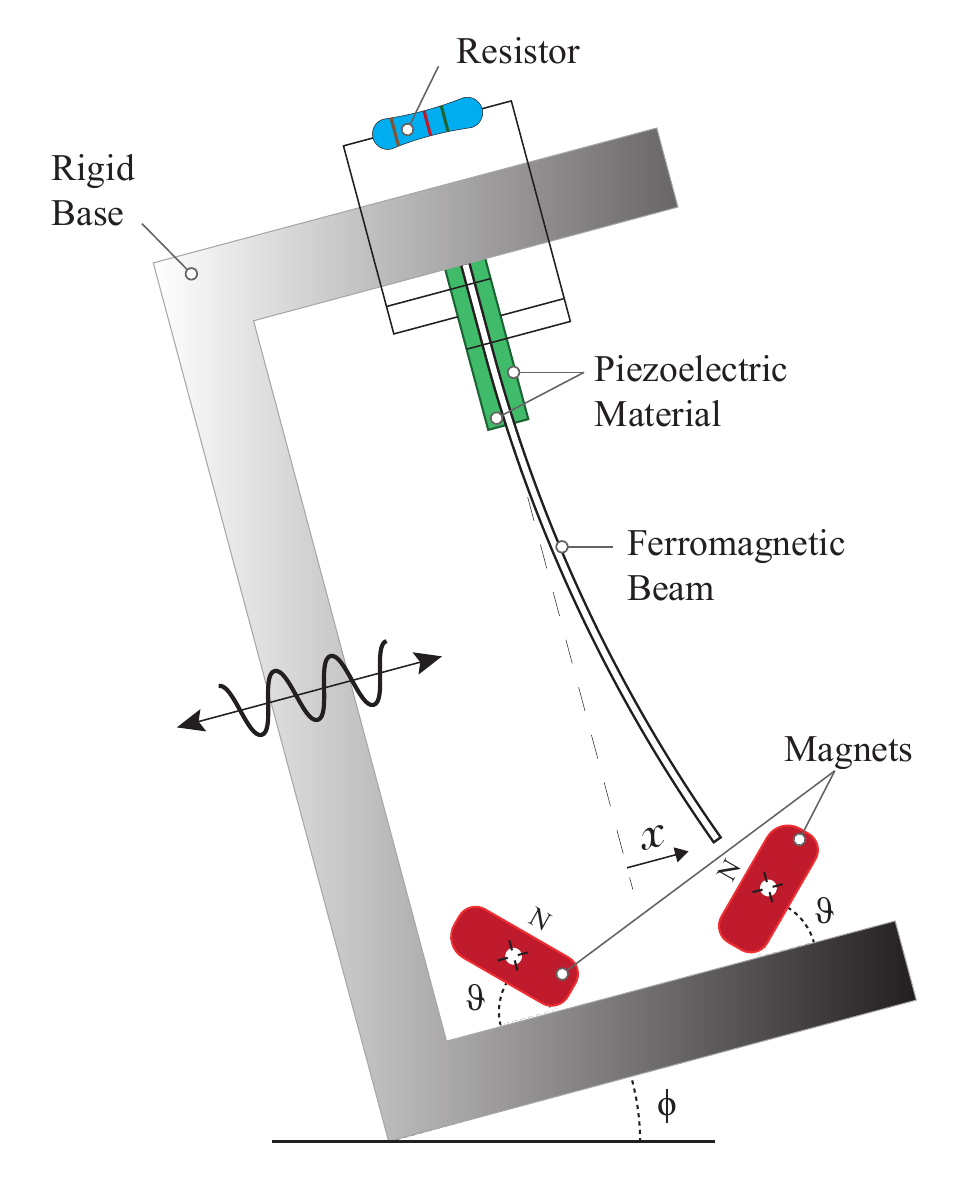}
    \caption{Schematic illustration of the Piezo-Magneto-Elastic beam energy harvester.}
    \label{fig:pmeh}
\end{figure}

An illustration of the energy harvesting system of interest in this work can be seen in Fig.~\ref{fig:pmeh}, which consists of a rigid base excited by a harmonic force, a vertical fixed-free beam made of ferromagnetic material, with two permanent magnets located near the bottom part of the base. In the high deformation region of the beam's upper part, a pair of piezoelectric layers is coupled to a resistive circuit responsible for converting the kinetic energy into an electrical potential, which is dissipated in the resistor. The harvester has a bias angle, leading to oscillations around a certain angle $\phi$. Parameter $\vartheta$ can be adjusted for different configurations in order to produce asymmetric potential energy functions.

Disregarding the asymmetric terms, the system is reduced to the classic piezo-magneto-elastic beam energy harvester proposed by Erturk, Hoffmann, and Inman \cite{8_Erturk}. Besides, this classic model uses a linear relationship between mechanical and electrical domains. Then, classic bistable energy harvester electromechanical equations are stated as
\begin{eqnarray}
 \ddot{\mathnormal{x}}+2\;\xi\;\dot{\mathnormal{x}}-\frac{1}{2}\;\mathnormal{x}\;(1-\mathnormal{x}^2)-\chi\; \mathnormal{v}  = \mathnormal{f}\;\cos{\left(\Omega\; t\right)} \, ,    \\
 \dot{\mathnormal{v}} + \lambda\;\mathnormal{v}+\kappa\; \dot x = 0 \, ,
    \\
\mathnormal{x}(0) = \mathnormal{x}_0, \;\dot{\mathnormal{x}}(0) = \dot{\mathnormal{x}}_0, \; \mathnormal{v}(0) = \mathnormal{v}_0 \, ,
\end{eqnarray}
where $\mathnormal{x}$ is the beam tip displacement; $\xi$ is the damping ratio; $\chi$ is the piezoelectric coupling term in the mechanical equation (depending on the piezoceramic properties and the structure mass); $\mathnormal{v}$ is the voltage; $\lambda$ is a reciprocal time constant ($\lambda\propto1/R_lC_p$, where $R_l$ is the load resistance and $C_p$ is the equivalent capacitance of the piezoceramic layers); $\kappa$ is the piezoelectric coupling term in electrical equation (depending on the piezoceramic properties and the equivalent piezoelectric capacitance); $f$ is the external excitation amplitude; $\Omega$ is the external excitation frequency. The initial conditions are $\mathnormal{x}_0$, $\dot{\mathnormal{x}}_0$ and $\mathnormal{v}_0$, which, respectively, represent the initial values of the beam edge position, velocity, and voltage through the resistor. All these quantities are dimensionless.

The main quantity of interest (QoI) associated with the dynamical system under analysis is the mean output power given by
\begin{equation}
     P \;=\;\frac1T\int_{t_0}^{t_0 + T} \lambda \;\mathnormal{v}(t)^2\;dt \, ,
     \label{eq:qio}
\end{equation}
which is the temporal average of the instantaneous power $\lambda\;\mathnormal{v}(t)^2$ over a given time interval of length $T$.

There are sources of nonlinearities that are unavoidable when building an energy harvester, for instance, the symmetrical position of the magnets. A quadratic nonlinearity is introduced to the nonlinear restoring force to characterize the asymmetric potentials \cite{Halvorsen,He_asymmetric}. Then, the restoring force is described in linear, quadratic, and cubic terms.

Another source of asymmetry often encountered in actual energy harvesters is the sloping plane where the system is attached, known as bias angle $\phi$. In many energy harvesting applications, the system is subjected to rotational movements that cause dynamic asymmetry. Because of the action of gravity, this system suffers changes in its inertial configurations. The resulting bias angle is used in \cite{Wang_enhancement} to enhance the harvester's performance so that it is added to the dynamic model as an external force from the self-weight of the harvester.

The nonlinear effect between the material strain and the piezoelectric coefficient can be seen in piezoceramic material according to \cite{11_Crawley}. In high deformation conditions, e.g., resonance regions, ignoring the nonlinear effects may cause an underestimation of harvested power \cite{12_duToit}. Therefore, models and studies considering a nonlinear electromechanical coupling for power harvesting are explored. Triplett and Quinn \cite{13_triplett} modeled the piezoelectric coupling by a first-order nonlinear relationship between material deformation and piezoelectric coefficient. In contrast, \cite{15_Leadenham,14_Stanton} also included higher-order terms on the piezoelectric constitutive equation.

In this way, to obtain more accurate estimates of the energy harvested by the system, it is also appropriate to consider a nonlinear electromechanical coupling. According to Triplett and Quinn \cite{13_triplett} the nonlinear coupling is determined through a dependent function on the linear coupling term $\theta_{L}$, nonlinear coupling term $\theta_{NL}$, and the deformation of the piezoelectric material ($s$). It follows that, the dimensional piezoelectric coefficient, obtained by \cite{13_triplett}, which is represented as
\begin{equation}
    \hat{\Theta}(\mathnormal{s}) = 
    \theta_{L} \, \left(1+\theta_{NL} \left|\mathnormal{s}\right| \right) \, .
\end{equation}

The following initial value problem describes the governing lumped-parameter equation of motion for the proposed bistable energy harvesting model taking into account asymmetries and nonlinear electromechanical coupling
\begin{eqnarray}
\begin{split}
 \ddot{\mathnormal{x}}+2\;\xi\;\dot{\mathnormal{x}}-\frac{1}{2}\;\mathnormal{x}\;(1+2\delta\mathnormal{x}-\mathnormal{x}^2)-(1+\beta\left|x\right|)\chi\; \mathnormal{v} \\  = \mathnormal{f}\;\cos{\left(\Omega\; t\right)}  + \mathnormal{p}\sin{\phi} \, ,
\end{split}
    \\
 \dot{\mathnormal{v}} + \lambda\;\mathnormal{v}+(1+\beta\left|x\right|)\kappa\; \dot x = 0 \, ,
    \\
\mathnormal{x}(0) = \mathnormal{x}_0, \;\dot{\mathnormal{x}}(0) = \dot{\mathnormal{x}}_0, \; \mathnormal{v}(0) = \mathnormal{v}_0 \, ,
\end{eqnarray}
where $\delta$ is a coefficient of the quadratic nonlinearity, $\mathnormal{p}$ is the equivalent dimensionless constant of gravity of ferromagnetic beam, and $\beta$ is the dimensionless nonlinear coupling term. 

\begin{figure*}
    \centering
    \includegraphics[width=0.95\textwidth]{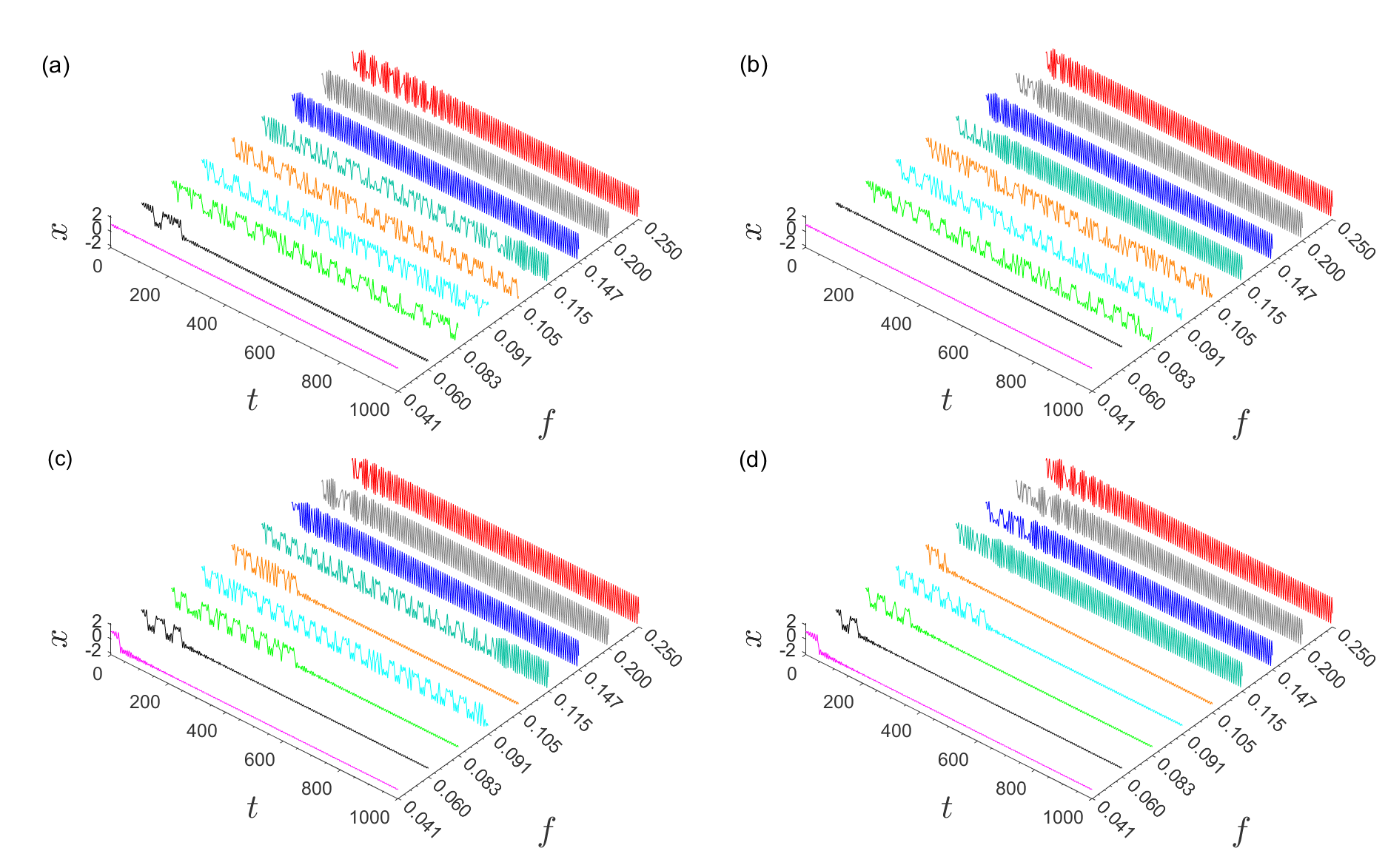}
    \caption{The displacement time-series for different amplitude of excitation. The time-series in (a) is classical bistable energy harvester ($\beta = \delta = \phi = 0$), in (b) classical bistable energy harvester with nonlinear coupling ($\beta = 1.0$ and $\delta = \phi = 0$), in (c) asymmetric bistable energy harvester ($\delta = 0.15$, $\phi = -10^{\circ}$ and $\beta = 0$), and in (d) asymmetric bistable energy harvester with nonlinear coupling ($\delta = 0.15$, $\phi = -10^{\circ}$ and $\beta = 1.0$).}
    \label{fig:time_series}
\end{figure*}

Figure~\ref{fig:time_series} depicts the displacement time-series for the classical bistable oscillators, with nonlinear electromechanical coupling, asymmetric potential, and bias angle under different amplitudes of excitation. These bistable oscillators have three equilibrium points: two stable (central region of each magnet) and one unstable (between the magnets). Thus, it presents three distinct dynamic behaviors, which mainly depend on the excitation amplitude. The lower energy orbits, i.e., oscillations only at one stable equilibrium point, are known for single-well motion. On the other side, the high-energy orbits, where the system oscillates between the equilibrium points (inter-well motion/ snap-through behavior), can be characterized as periodic and chaotic behaviors. In energy harvesting, vibrations at high-energy orbits can improve the system's performance. This figure also displays that piezoelectric nonlinearities, asymmetric geometry, and bias angle can change the dynamic behavior compared to the classical case, mainly in the unstable region with a low excitation amplitude. In addition, the high sensitivity of this type of system is observed. An animation of these systems is available in the Supplementary Material and also in the STONEHENGE code repository \cite{STONEHENGE}, which is shown in detail each behavior for these systems over time. The playlist at \cite{STONEHENGE} also presents the linear and nonlinear harvesters' dynamic animations.

\section{Global sensitivity analysis}
\label{sec:Global sensitivity}

This section presents the global sensitivity analysis framework to study the bistable energy harvester dynamics. Relevant insights are provided into the system in question that allows model reduction schemes; for instance, it builds a simplified model by fixing relatively unimportant parameters to their nominal values and varying only the most influential parameters \cite{Sudret2020PCA}. The global sensitivity analysis method adopted is a variance decomposition that writes the dynamical system response as a sum of each input variable's individual and combined contributions. The Sobol indices, established in \cite{29_Sobol} quantifies the contribution of each parameter concerning the total variance of the model. It has recently been used in many works, with high impact research, as in \cite{Alemazkoor,Lund}. One of its main advantages is dealing with nonlinear and non-parameterized models and providing a quantitative and qualitative classification.

To facilitate mathematical development, consider that the QoI from the dynamical system of interest can be represented by the following functional relationship
\begin{equation}
     Y = \cal{M}(\textbf{X})  ~~~\mbox{,}~~~
	        \textbf{X} = \{\mathnormal X_{\mathnormal{1}},\mathnormal X_{\mathnormal{2}},\dots,\mathnormal X_{\mathnormal{k}} \},
\end{equation}
where $\textbf{X}$ is an input vector with $k$ independent parameters, which is modified by the mathematical operator  $\cal{M}$,  to produce the scalar output (the QoI) $Y$. 

Using the Hoeffding-Sobol decomposition \cite{Hoeffding,29_Sobol}, the QoI can be written as a decomposition into summations of different dimensions
\begin{eqnarray}
    Y=\cal{M}_\mathnormal{0} + 
    \sum_{\mathnormal{i}=\mathnormal{1}}^\mathnormal{k} \cal{M}_\mathnormal{i}\left(\mathnormal X_\mathnormal{i}\right) +
    \sum_{\mathnormal{i}<\mathnormal{j}} \cal{M} _{\mathnormal{i}\mathnormal{j}}\left(\mathnormal X_\mathnormal{i},\mathnormal X_\mathnormal{j}\right) + \nonumber \\ \dots +
    \cal{M}_{\mathnormal{1}\dots\mathnormal{k}}\left(\mathnormal X_{\mathnormal{1}} \dots \mathnormal X_\mathnormal{k}\right) \, ,
    \label{eq:sobol_decom}
\end{eqnarray}
\noindent
which has finitely many sums with finite variance, and according to Hommaa and Saltelli \cite{30_Homma} it has an orthogonal decomposition property in terms of conditional expectations
\begin{equation}
    \int_{{\cal D}_x}{\cal M}_u\left(x_u\right){\cal M}_v\left(x_v\right) \mathnormal{d}x\;=\;0, ~~~~~~ \forall~ u\neq v
\end{equation}
where $u \subset\left\{1,\dots,k\right\} $ and $x_u {\stackrel{def}{=}}\left\{x_{i\mathnormal{1}},\dots,x_{is}\right\}$.

In order to simplify the notation, the input variables are assumed to follow uniform distributions with a normalized support $[0, 1]$, so that the support of the random input vector is ${\cal D}_X = [0, 1]^k$. Note that it is always possible to normalize a finite support random variable. Thus, all the sums in the expansion can be recursively computed. The first term $\cal{M}_\mathnormal{0}$ is a constant equal to the expected value of $Y$, i.e.,
\begin{equation}
     \cal{M}_\mathnormal{0} = \int_{{\cal D}_\mathnormal{X}}\cal{M}\left(\textbf{X}\right) \mathnormal{d} \textbf{X} \, .
\end{equation}

The $\cal{M}_\mathnormal{i}(\mathnormal X_{\mathnormal{i}})$ and $\cal{M}_{\mathnormal{i}\mathnormal{j}}(\mathnormal X_{\mathnormal{i}},\mathnormal X_{\mathnormal{j}})$ terms are, respectively, the conditional expected value of $\cal{M}$ given the parameters $\mathnormal{i}$ and combinations of the parameters $\mathnormal{ij}$ ($\mathnormal{i}\neq\mathnormal{j}$), written as
\begin{equation}
     \cal{M}_\mathnormal{i}(\mathnormal X_{\mathnormal{i}}) = \int_\mathnormal{0}^\mathnormal{1}\dots\int_\mathnormal{0}^\mathnormal{1} \cal{M}\left(\textbf{X}\right) \mathnormal{d}\textbf{X}_{\sim \mathnormal{i}} - \cal{M}_\mathnormal{0} \, ,
\end{equation}

\begin{eqnarray}
     \cal{M}_{\mathnormal{i}\mathnormal{j}}(\mathnormal X_{\mathnormal{i}},\mathnormal X_{\mathnormal{j}}) = \int_\mathnormal{0}^\mathnormal{1}\dots\int_\mathnormal{0}^\mathnormal{1} \cal{M}\left(\textbf{X}\right) \mathnormal{d} \textbf{X}_{\sim \mathnormal{ij}} - \cal{M}_\mathnormal{0} - \nonumber\\ \cal{M}_\mathnormal{i}(\mathnormal X_{\mathnormal{i}}) - \cal{M}_\mathnormal{j}(\mathnormal X_{\mathnormal{j}}) \, ,
\end{eqnarray}
where the notation $\sim$ indicates which variables are excluded from the integration.

This allows the meaningfully partition the system response variance and determine the Hoeffding-Sobol decomposition in terms of variance \cite{29_Sobol}. Considering independent input variables, we can quantify the contribution of the individual $ \mathnormal{X}_\mathnormal{i}$ and any combination $\mathnormal{X}_{\mathnormal{ij}},\dots,\mathnormal X_{\mathnormal{i}\dots\mathnormal{k}}$ of variables to the total variance $Var[Y]$. By construction
\begin{equation}
    \sum_{\mathnormal{i}=1}^\mathnormal{k} S_\mathnormal{i} +     \sum_{\mathnormal{i}<\mathnormal{j}} S_{\mathnormal{i}\mathnormal{j}} + \dots + S_{\mathnormal{1 2} \dots \mathnormal{k}} = 1 \, ,
    \label{eq:sobol_sum}
\end{equation}
where the indices $S$ are called as Sobol indices and the sum of all individual and joint indices is equal to the unity.

The first-order Sobol indices
\begin{equation}
    S_{\mathnormal{i}} = \frac{Var[\cal{M}_{\mathnormal{i}}(\mathnormal X_{\mathnormal{i}})]}{Var[\cal{M}(\textbf{X})]} \, ,
\end{equation}
quantify the additive effect of each input separately concerning the total variance, and the second-order Sobol indices $S_{{\mathnormal{i}}{\mathnormal{j}}}$
\begin{equation}
    S_{{\mathnormal{i}}{\mathnormal{j}}} = \frac{Var[\cal{M}_{{\mathnormal{i}}{\mathnormal{j}}}(\mathnormal X_{\mathnormal{i}},\mathnormal X_{\mathnormal{j}})]}{Var[\cal{M}(\textbf{X})]} \, ,
\end{equation}
compute the joint-effects of two inputs. Higher-order Sobol indices consider the interaction effects of several parameters and are written in the same way.

\section{Polynomial chaos expansion}
\label{sec:PCE}

Generally, Sobol indices are calculated using the Monte Carlo (MC) method, but this can be computationally expensive due to the slow convergence rate of this sampling method \cite{CunhaJR_MC,Kroese_MC}. An alternative way to obtain Sobol indices is to employ the Polynomial Chaos Expansion (PCE) surrogate model, which is an approximate model for the original system with a short processing time, reducing the computational cost and keeping calculation accuracy \cite{37_Crestaux,38_Palar,31_sudret}.

PCE was introduced in the engineering community by R. Ghanem, and P. Spanos \cite{33_Ghanem} in the context of stochastic finite element analysis. The PCE is used for UQ in various domains, e.g., in solid mechanics, thermal and fluid sciences, etc. It can be a non-intrusive method representing uncertain quantities as an expansion, including the decomposition of deterministic coefficients and orthogonal polynomial bases for each random variable probability distribution \cite{36_Xiu}. PCE offers an accurate and efficient way to include nonlinear effects in stochastic analysis and can be seen as an ideal mathematical way to construct and obtain a model response surface in the form of a high-dimensional polynomial in model parameters uncertain \cite{34_Oladyshkin,35_SEPAHVAND}.

Assuming that $Y = \cal{M}(\textbf{X})$ is a finite variance random variable, defined in terms of the composition of the random vector $\textbf{X}$ and the operator $\cal{M}$, one can write the following PCE
\begin{equation}
    Y \approx 
    \sum_{\alpha \in \cal{A}} \mathnormal{y}_\alpha\psi_\alpha(\textbf{X}) \, ,
\end{equation}
where, $\psi_\alpha$ are multivariate orthonormal polynomials  concerning the joint probability density function (PDF) $f_X$ of $\textbf{X}$, $\mathnormal{y}_\alpha$ are unknown deterministic coefficients, and the truncation set $\cal{A} \subset \mathbb{N}^\mathnormal{M}$ is selected from all possible multi-indices of multivariate polynomials. The unknown coefficients can be obtained from a non-intrusive least-squares regression technique, where samples from the system response are obtained from the full dynamics model.

Statistical moments and sensitivity analysis can be computed efficiently via post-processing of the PCE coefficients. Due to orthonormality property of the PCE basis, and the fact that $\psi_0 \equiv 1$ and $\mathbb{E}~[\psi_\alpha(\textbf{X})] = 0 ~\forall~\alpha~\neq 0$, the  mean value and the variance of the system response can estimated as
\begin{equation}
    \mathbb{E}\left[Y\right] \approx \mathnormal{y}_0 \, ,
\end{equation}
and
\begin{equation}
    \mathbb{E} \left[\left(Y- \mathbb{E}\left[ Y \right]\right)^2\right] \approx \sum_{\stackrel{\alpha\neq0}{\alpha \in \cal A}}\mathnormal{y}_\alpha^2 \, .
\end{equation}

Similarly, the Sobol indices of any order can also be determined analytically, e.g., the first-order indices are derived as follows
\begin{equation}
    S_i \approx \sum_{
    \stackrel{\alpha\neq0}{\alpha \in \cal{A}_\mathnormal{i}}} \mathnormal{y}_\alpha^2\bigg/\sum_{\stackrel{\alpha\neq0}{\alpha \in \cal A}}\mathnormal{y}_\alpha^2 \, ,
\end{equation}
where the PCE coefficients for the numerator are restricted to the subset $\cal{A}_{\mathnormal{i}}$, according to the parameter $\mathnormal{i}$, to compute the conditional variance. The right side of the equation is an approximation due to the truncation scheme used to compute the PCE. Similarly, the second-order Sobol indices are stated as
\begin{equation}
    S_{ij} \approx \sum_{\stackrel{\alpha\neq0}{\alpha \in \cal{A}_\mathnormal{ij}}}\mathnormal{y}_\alpha^2\bigg/\sum_{\stackrel{\alpha\neq0}{\alpha \in \cal A}}\mathnormal{y}_\alpha^2 \, .
\end{equation}

\begin{figure*}
\centering
\includegraphics[width=0.95\textwidth]{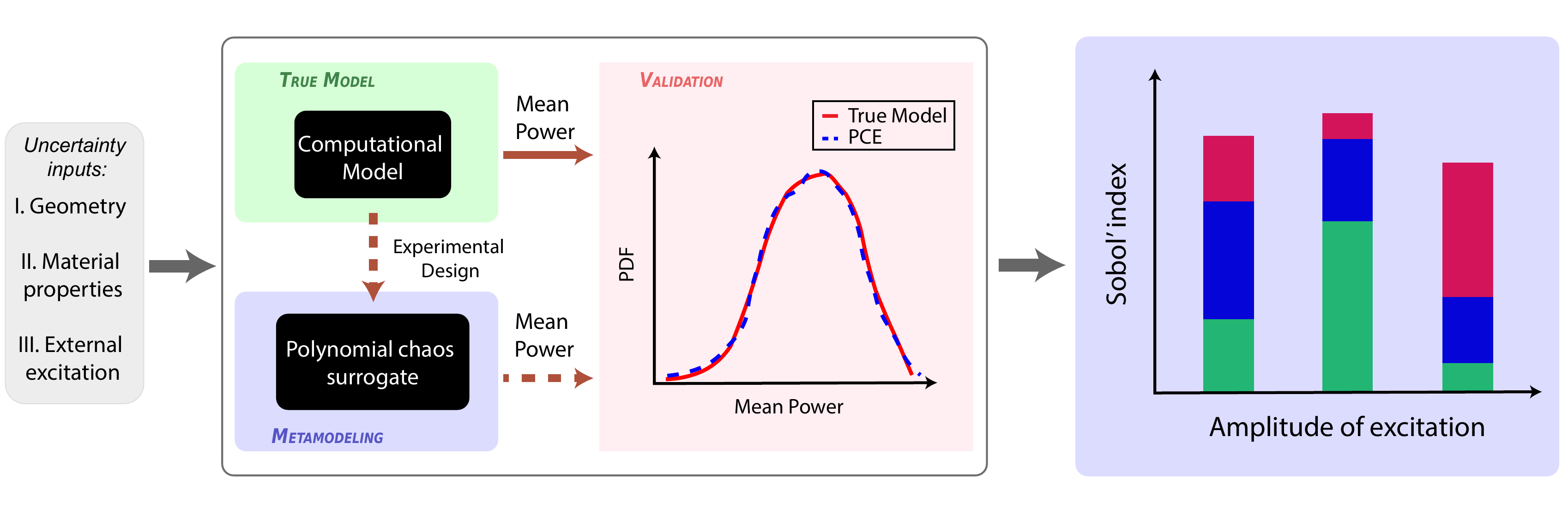}
\caption{Methodology schematic for the sensitivity analysis of the energy harvester about parameter variations.}
\label{fig:project}     
\end{figure*}

Figure~\ref{fig:project} outlines a summary of the methodology employed to assess the sensitivity of the system response regarding variations in the parameters.

\section{Results and discussion}
\label{sec:NE}

Numerical experiments reported in this work adopt the following physical parameters: $\xi = 0.01$, $\chi = 0.05$, $\lambda = 0.05$, $\kappa = 0.5$, $\Omega = 0.8$ and $f$ is varied. The initial condition is defined by $(\mathnormal{x}_0,\dot{\mathnormal{x}}_0,\mathnormal{v}_0) = (1,0,0)$. The dynamics is integrated by Runge Kutta scheme (4th order) over the time interval $0 \leq t \leq 2000$ with relative tolerance of $10^{-6}$, and absolute tolerance of $10^{-9}$. The mean output power is computed over the last 50\% of these time-series. 

Global sensitivity analysis assumes that all the system parameters are independent and uniformly distributed over given intervals, defined by a coefficient of variation ($c.v.$) of $\pm$~20\% around the nominal values.

We approached different models to verify specific behaviors for each condition: The classical model without nonlinear electromechanical coupling and asymmetries is evaluated. In the following, the nonlinear electromechanical coupling is included. The asymmetry bistable model with linear piezoelectric is studied. Finally, a proposed model with nonlinear coupling and asymmetries is approached.

\subsection{Classical bistable energy harvester}

\begin{figure}
    \includegraphics[width=0.98\textwidth]{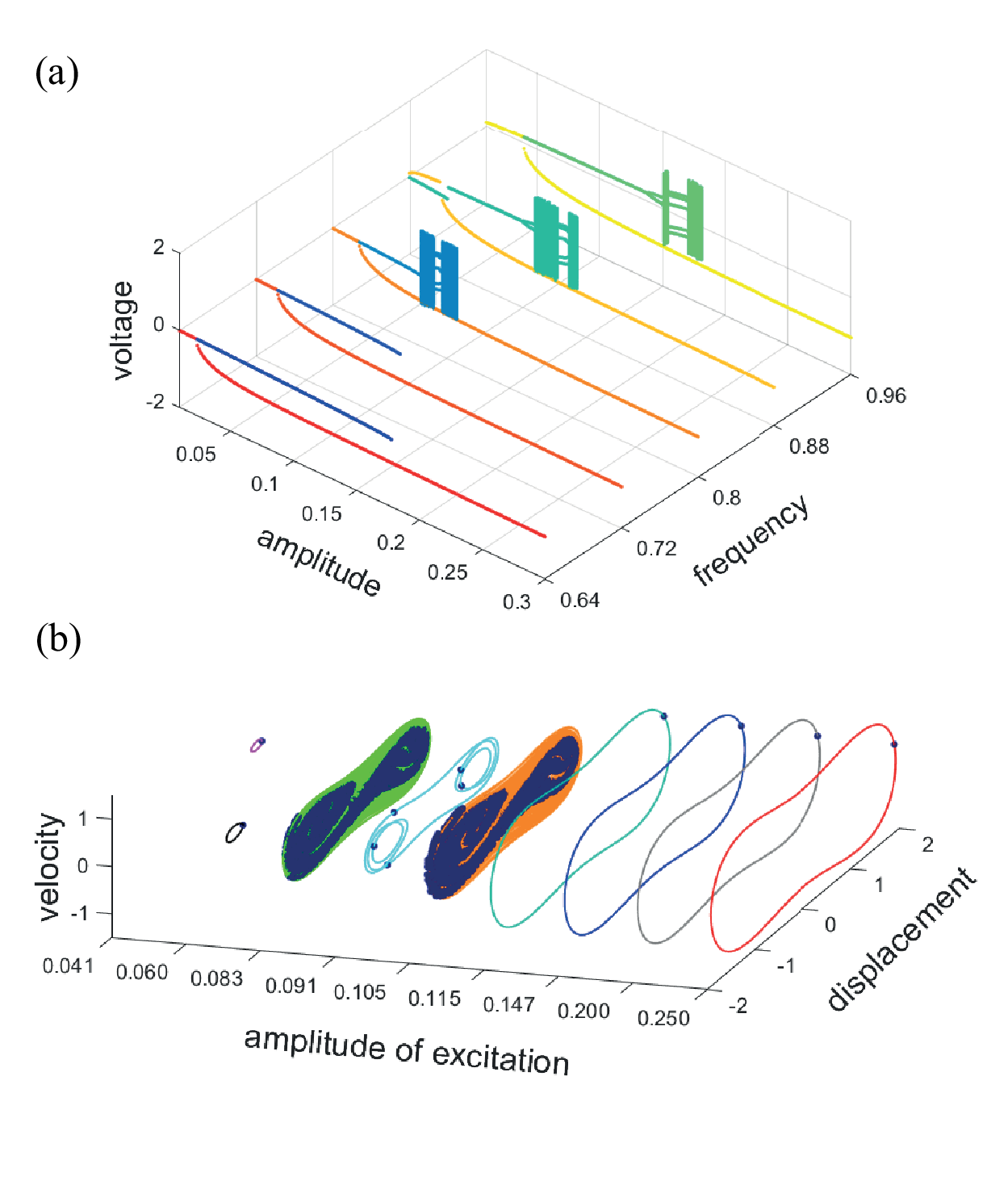}
    \caption{Numerical simulations results for the bistable energy harvester with linear electromechanical coupling: (a) bifurcation diagrams of voltage as a function of the excitation amplitude several values of the excitation frequency. The sweeping up the amplitude excitations are presented with cool colors while sweeping down appears in hot colors. (b) Phase portrait and Poincare section for different excitation amplitudes and $\Omega = 0.8$.}
    \label{fig:bifurc_pmeh}      
\end{figure}

The classical bistable oscillator is sensitive to excitation conditions as shown in \cite{Lopes}. So, the system's behavior through bifurcation diagrams with nominal parameters is assessed. The goal is to identify regions of the amplitude and frequency of excitation that imply distinct nominal dynamic behavior, being an essential step in the sensitivity analysis. Figure~\ref{fig:bifurc_pmeh}a shows the bifurcation diagrams of the classical bistable energy harvester. In the bifurcation diagram, sweeping the excitation amplitude to five frequency values established within the variance range is performed. For excitation frequencies of 0.64 and 0.72, the system presents a periodic behavior for the entire amplitude range. It is possible to identify low-energy orbit regions for lower amplitude values and high-energy orbits for larger values through a phase portrait and Poincare section in Fig.~\ref{fig:bifurc_pmeh}b. Alternatively, excitation frequencies of 0.8, 0.88, and 0.96 can lead to chaotic behavior and inter-well motion.

Two different dynamic behaviors from the bifurcation diagram are chosen to validate the metamodel, testing distinct scenarios. The first-order Sobol indices are calculated using the Monte Carlo method. The surrogate PCE model is constructed and tuned using Monte Carlo simulations as a reference. It is worth highlighting that the PCE model requires verification from a reference to guarantee the reliability of convergence and accuracy, as shown in the schematic representation of the methodology, Fig.~\ref{fig:project}. Finally, the accurately PCE model is used to explore several scenarios of excitation amplitude with low-cost processing. To illustrate this process, is selected nominal values $f = 0.147$ and $\Omega = 0.8$, as a periodic behavior and $f = 0.083$ and $\Omega = 0.8$, as a chaotic one.

\begin{figure*}
    \includegraphics[width=0.94\textwidth]{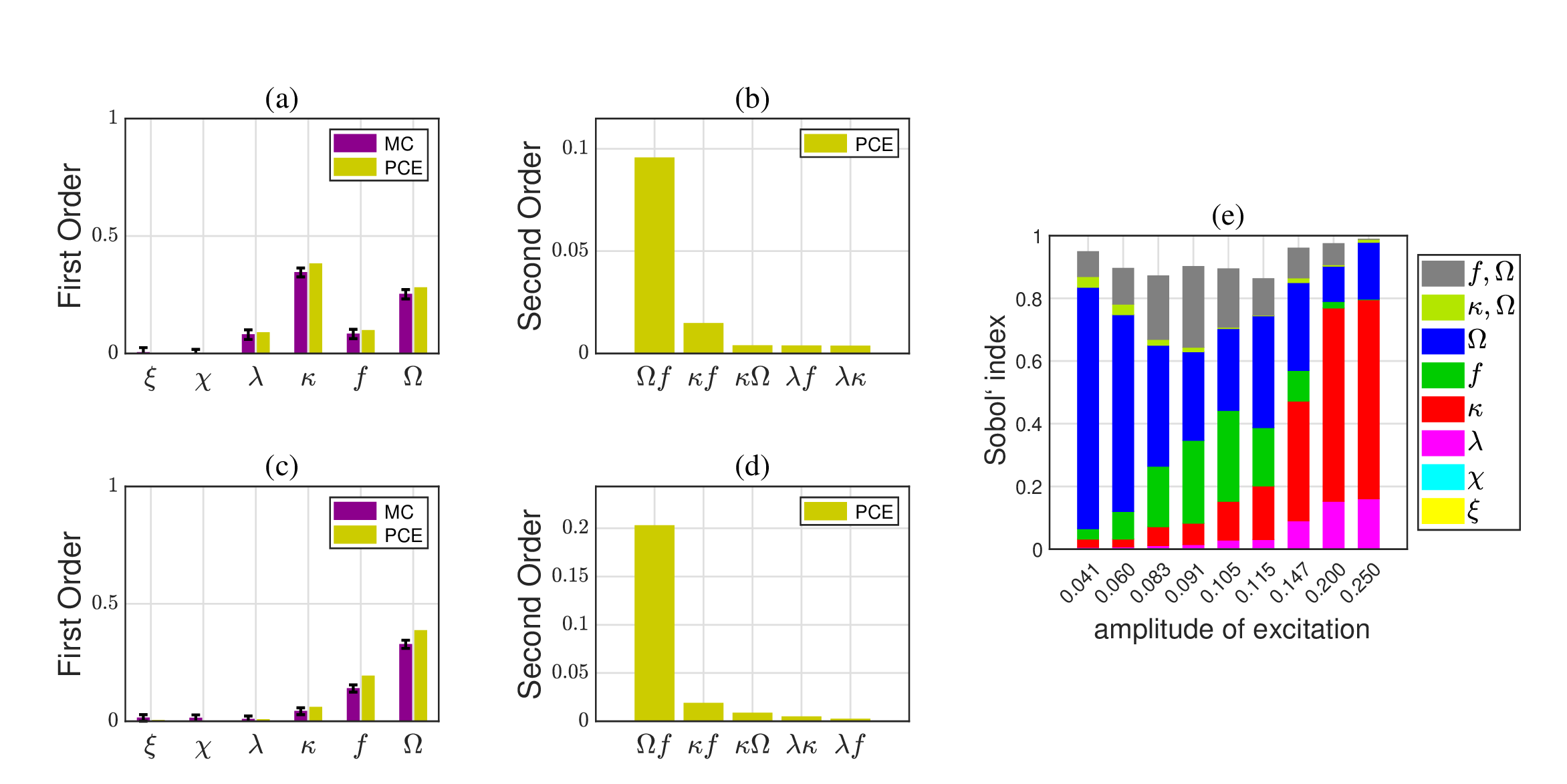}
    \caption{Sobol' indices associated with the mean output power for the bistable energy harvester with linear coupling. First for nominal regular steady-state dynamics ($f = 0.147$ and $\Omega = 0.8$): in \textbf{(a)} the first-order based on MC and PCE methods and in \textbf{(b)} the second-order based on PCE method. Second for nominal irregular steady-state dynamics ($f = 0.083$ and $\Omega = 0.8$): in \textbf{(c)} the first-order based on MC and PCE methods and in \textbf{(d)} the second-order based on PCE method. Finally, \textbf{(e)} presents the main Sobol' indices exploring several scenarios with different excitation amplitudes based on the PCE method.}
    \label{fig:sobol_rang}  
\end{figure*}

The first-order Sobol indices for the periodic case are shown in Fig.~\ref{fig:sobol_rang}a using MC and PCE methods. These indices reveal that $40\%$ of the system sensitivity is due to the piezoelectric coupling acting on the electrical equation $\kappa$, then $30\%$ for the frequency $\Omega$, $10\%$ for amplitude $f$, and $10\%$ for the reciprocal time constant $\lambda$. The influence of the piezoelectric coupling on the mechanical system $\chi$ and the damping ratio $\xi$ did not cause variability in power harvested. 

For a second case, dealing with nominal chaotic behavior, the first-order Sobol indices are shown in Fig.~\ref{fig:sobol_rang}c based on MC and PCE methods. For this scenario, the frequency and amplitude are powerful for energy-harvested variability, supplying about 40\% and 20\% of the total variance, respectively. The piezoelectric coupling on the electrical equation has 5\%; finally, the reciprocal time constant, piezoelectric coupling on the mechanical equation, and damping have negligible values.

To compare the employed methods, in Tab.~\ref{tab:tab1} shows the Sobol Indices performance between MC and PCE for the periodic and chaotic case. Note that PCE presents a shorter time processing than MC. The mean absolute errors (MAE) for Sobol Indices between the methods are low for both cases. Besides, the Supplementary Material presents the PCE approach by PDF compared to the complete model response, showing good performance. Then, PCE is accurate and can substitute precisely our model, becoming workable to explore several behaviors and high orders of Sobol indices.

\begin{table}[ht]
\centering
\caption{Comparison of the computational cost of the Sobol indices via MC and PCE methods.}
\begin{tabular}{ccccc}
\specialrule{.07em}{.05em}{.05em}\noalign{\smallskip}
\textbf{case} & \textbf{method} & \textbf{\begin{tabular}[c]{@{}c@{}}sample \\ size\end{tabular}}  & \textbf{\begin{tabular}[c]{@{}c@{}}CPU time* \\ (seconds)\end{tabular}} & \textbf{MAE} \\
\noalign{\smallskip}\specialrule{.07em}{.05em}{.05em}\noalign{\smallskip}
\multirow{2}{*}{\textit{periodic}} & MC & 10~000 & $\approx$ 86~400 & \multirow{2}{*}{1.59\%} \\ 
 & PCE & 1~000 & $\approx$ 2~700 & \\ \noalign{\smallskip} \hline \noalign{\smallskip}
\multirow{2}{*}{\textit{chaotic}}  & MC & 20~000 & $\approx$ 162~000 & \multirow{2}{*}{2.57\%} \\
 & PCE & 2~000 & $\approx$ 6~000  & \\
\noalign{\smallskip}\specialrule{.07em}{.05em}{.05em}
\end{tabular}
\\ \smallskip
\textbf{*}\textsf{Intel i7-9750H 2.60GHz 8GB 2666GHz DDR4}
\label{tab:tab1} 
\end{table}

Figure~\ref{fig:sobol_rang}b presents the second-order indices for the first case, where they are calculated only by the validated PCE method. This result shows that only the combined effect of the excitation parameters is relevant, meaning about 10\% remaining of the total variance of the system. For higher orders, their evaluations are unnecessary since they have accounted for (almost) the distribution of 100\% of the total variance. The second-order Sobol indices for the second case are shown in Fig.~\ref{fig:sobol_rang}d and demonstrate that only the joint effect of excitation parameters $\Omega f$ is relevant, about 20\%. The increase of the $\Omega$ and $f$ effects occur since chaotic behavior is more sensitive to excitation conditions, as proven in first-order Sobol indices.

A wide range of excitation amplitudes using Sobol indices based PCE is explored. Sobol indices are calculated for each excitation amplitude. The indices are presented by a stacked bar plot in Fig.~\ref{fig:sobol_rang}e. All the first-order indices and the relevant second-order indices are inserted. The $\Omega$, $f$, and $\kappa$ effects are the parameters that most cause variability in harvested power because they present larger indices for all ranges. In regions with a wide range of stability, where the system has enough energy to oscillate in inter-well motion - high excitation amplitude - the sensitivity to $\kappa$ is the most relevant. However, the effects of $\Omega$ and $f$ increase in the unstable condition, where the system can change the dynamic behavior for a lower amplitude of excitation (as seen in Fig.~\ref{fig:bifurc_pmeh}). Finally, Fig.~\ref{fig:sobol_rang}e also shows that the higher-order terms have a negligible effect, because the first- and second-order indices account for almost all variance: the bars are reaching close to 1 (see Eq.~\ref{eq:sobol_sum}).

The effects of frequency and amplitude of excitation are significant because their variability can change dynamic behaviors, which is seen in the bifurcation diagram under the $ c.v. = \pm~20\% $ at high frequency. The dynamic behavior changes directly affect the power harvested, e.g., high-energy orbits produce much more electricity than chaotic or low-energy orbits. Although the frequency and amplitude of excitation are mainly responsible for behavior changes, the piezoelectric coupling presents greater sensitivity for high amplitude. This is due to the small probability of behavior variation, so the piezoelectric effect predominates. Therefore, whatever scenario adjacent to the orbit changes, $f$, and $\Omega$ are the most influential parameters; otherwise, $\kappa$ is more relevant.

\subsection{Classical bistable energy harvester with nonlinear electromechanical coupling}

\begin{figure}
\centering
\includegraphics[width=0.95\textwidth]{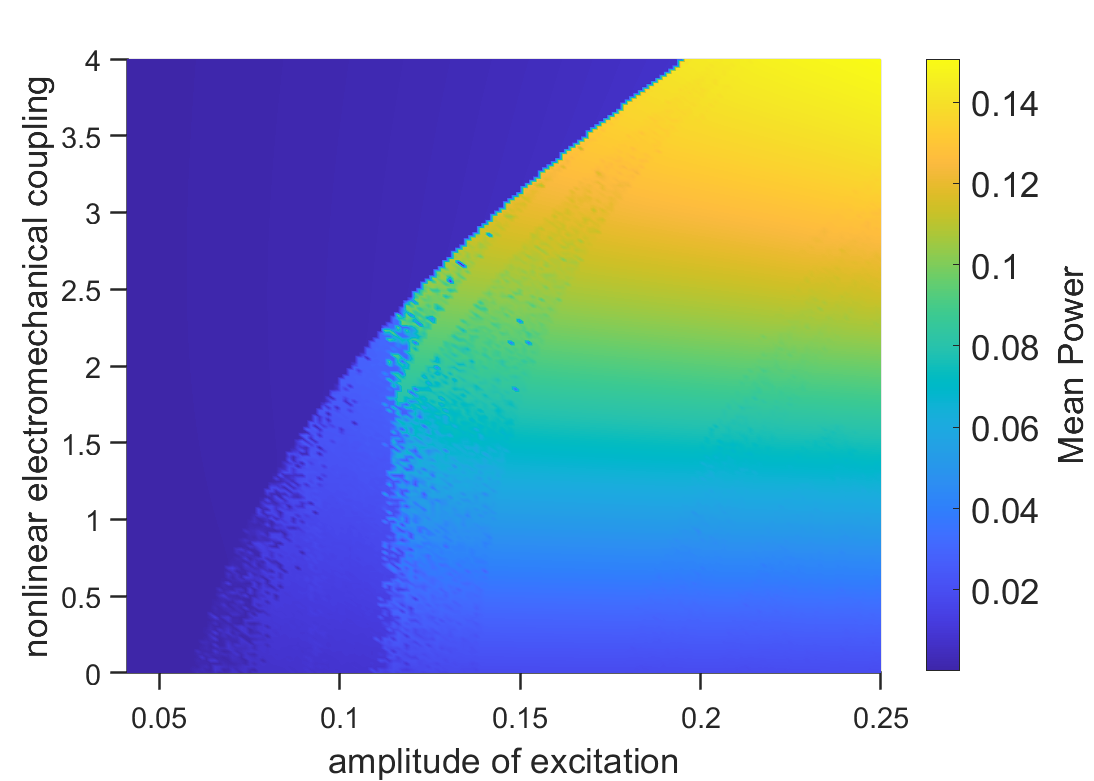}
\caption{Contour map of the mean output power over a wide range of excitation amplitude and nonlinear electromechanical coupling.}
\label{fig:surf_Brang} 
\end{figure}

The nonlinear electromechanical coupling is added in the classical bistable energy harvester. First, a parametric investigation of nonlinear electromechanical coupling term $\beta$ and amplitude excitation range through the mean power performance is performed. Fig.~\ref{fig:surf_Brang} shows a contour map for the mean power over a wide range for the excitation amplitude and the nonlinear electromechanical coupling. The system can undergo an increase in the harvested power as $\beta$ increases. However, there is a threshold value for the nonlinear term, damaging system performance from that point on. The high values of the nonlinear coupling term harm the power harvested for low excitation amplitude. Besides, in instability regions (chaotic behavior), the system has more significant mean power harvested changes, noticing a discontinuity color region between $f = 0.07$ to $0.105$ (chaotic area). This situation exemplifies how the nonlinear electromechanical coupling requires careful investigation, leading to a more complex analysis.

\begin{figure}[!htb]
    \centering
    \includegraphics[width=0.95\textwidth]{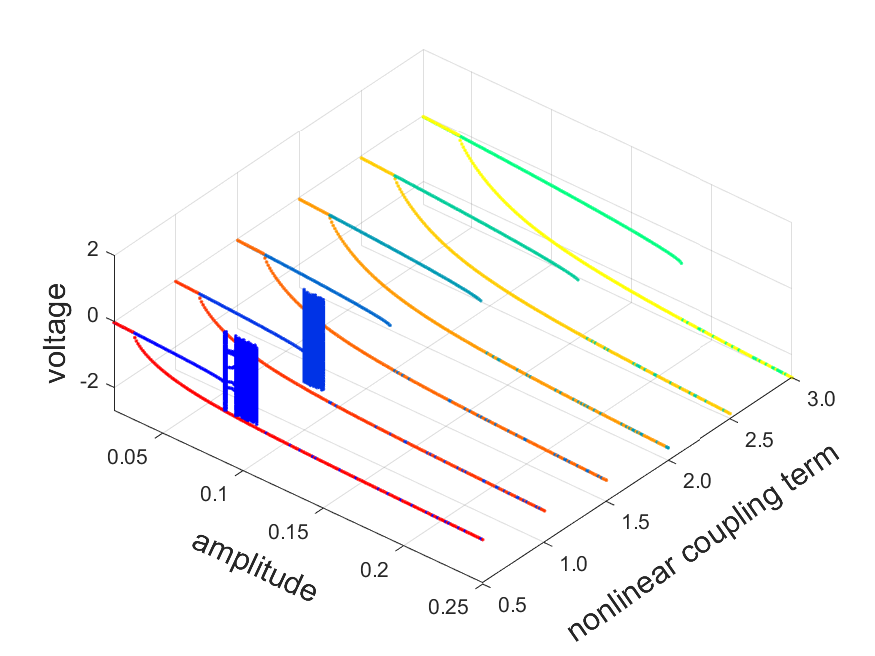}
    \caption{Bifurcation diagrams of voltage for the bistable energy harvester with nonlinear electromechanical coupling as a function of the excitation amplitude, for six values of nonlinear coupling term: $\beta \in\{ 0.5, 1.0, 1.5, 2.0, 2.5, 3.0\}$ and $\Omega = 0.8$. The sweeping up the amplitude excitation is presented with cool colors, while sweeping down appears in hot colors.}
    \label{fig:bifurc_pmehn}
\end{figure}

Figure~\ref{fig:bifurc_pmehn} shows the bifurcation diagram sweeping up and down the amplitude of excitation for different nonlinear coupling values ($\beta = 0.5, 1.0, 1.5, 2.0, 2.5$ and $3.0$). For low values of the nonlinear coupling, the system maintains almost the same behavior as the absolute amplitude. The nonlinear coupling only decreases the chaotic interval, shifting the chaos onset to higher excitation amplitudes. However, the regions with chaotic intervals vanish as $\beta$ increases, and the dynamic behavior acquires a large range of low-energy orbits, as seen in Fig.~\ref{fig:surf_Brang} with a large area of low energy as the nonlinear coupling increases (dark blue region). This effect occurs because the nonlinear coupling acts as damping in the system, as explained in \cite{13_triplett}. The nonlinear piezoelectric configuration harvests more energy in high-energy orbits, but it demands more external energy for the mechanical system to perform a snap-through behavior.

\begin{figure*}[!htb]
\includegraphics[width=0.95\textwidth]{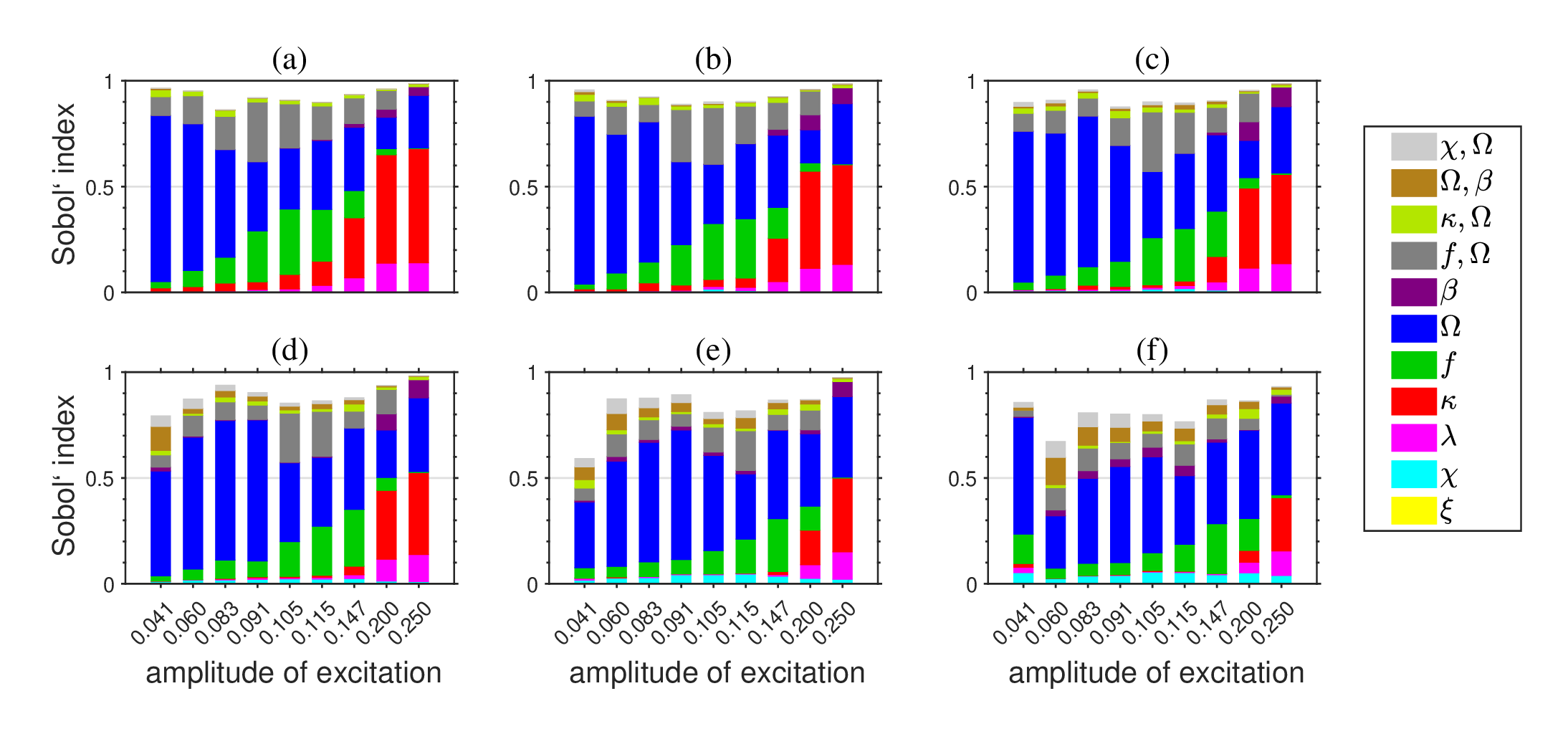}
\caption{Sobol indices based on PCE method associated with the mean output power of the bistable energy harvester for several scenarios of excitation amplitude with nonlinear coupling, taken into account: \textbf{(a)} $\beta = 0.5$, \textbf{(b)} $\beta = 1.0$, \textbf{(c)} $\beta = 1.5$, \textbf{(d)} $\beta = 2.0$, \textbf{(e)} $\beta = 2.5$, \textbf{(f)} $\beta = 3.0$.}
\label{fig:sobol_rang_pmehn}
\end{figure*}

In Fig.~\ref{fig:sobol_rang_pmehn}, the leading Sobol indices are shown for different $\beta$ values for several amplitudes excitation scenarios after obtaining a PCE surrogate in the same way as done in the classical case. More detail on PCE accuracy can be seen in the Supplementary Material. 

The sensitivity to the nonlinear piezoelectric term has a minor influence. However, the higher $\beta$ values ($\beta\geq2$) change the individual and combined effect of other parameters, i.e., $\beta$ indirectly affects the system's sensitivity. The effect of the excitation frequency is dominant throughout the amplitude spectrum. The excitation amplitude has a slight drop in influence, but its contribution is consistently significant in the chaotic regions for low values of $\beta$. The electromechanical coupling in the electrical equation also loses some influence, mainly because the nonlinear term drives the system more sensitive to excitation conditions. Higher $\beta$ values require more external energy to perform inter-well motions and achieve a stable region. Also, there is a significant increase in the combined influence of $\Omega \, \beta$ and $\xi \Omega$, previously insignificant. It concludes that when the piezoelectric nonlinearity is high, the system may change its dynamics, affecting its sensitivity and indirectly increasing and decreasing the individual and combined influence of other parameters. These results emphasize that information about nonlinear coupling is crucial. High values for the piezoelectric nonlinearity may change the dynamics, affecting its sensitivity and indirectly interfering with other parameters' individual and joint influence. On the other hand, it is unnecessary to worry about its uncertainties for low nonlinearity. Therefore, we recommend considering the uncertainties for high nonlinearities.

\subsection{Asymmetric bistable energy harvester with linear electromechanical coupling}

\begin{figure}
    \centering
    \includegraphics[width=0.85\textwidth]{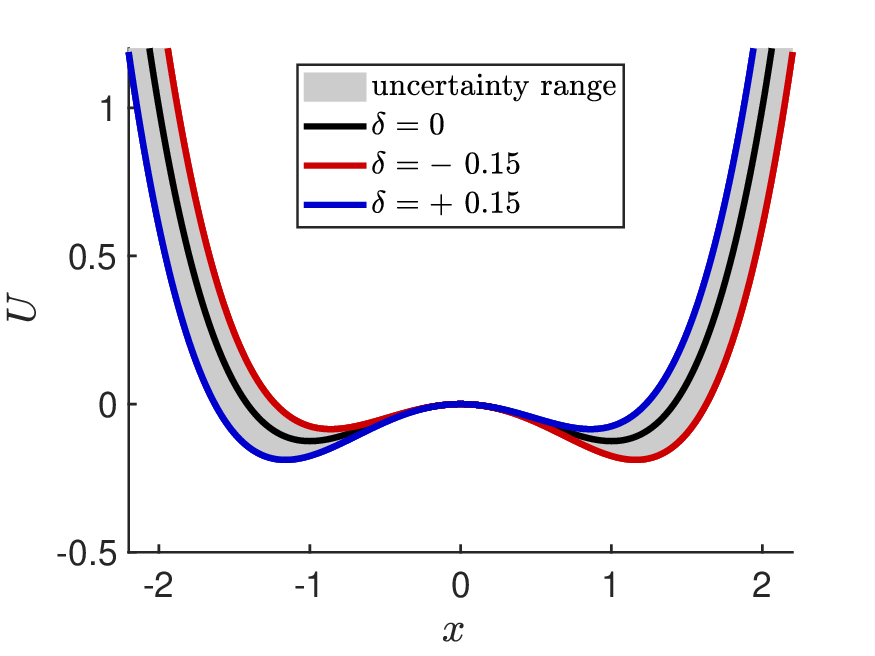}
    \caption{Illustration of the potential energy function, of symmetric and asymmetric configurations, for $\delta$ equals 0.15 and -0.15. Furthermore, an uncertainty range of $\delta$ is considered.}
    \label{fig:Potential_energy}
\end{figure}

This section treats the classical bistable energy harvesting system considering asymmetries. The asymmetric potential, as previously discussed, is defined by a quadratic term, but its coefficient depends on the different variations of each design. The asymmetric term is considered through a random variable described by a uniform distribution ranging from $-0.15$ to  $0.15$. Figure~\ref{fig:Potential_energy} shows the potential energy of the system for different $\delta$ values. The $\delta$ changes the equilibrium point of the model, causing asymmetry in the potential. Left asymmetry occurs for positive values and right asymmetry for negative values. Then, the sensitivity of the system considering the energy potential varying in this uncertainty range is analyzed.

\begin{figure}
    \centering
    \includegraphics[scale=0.36]{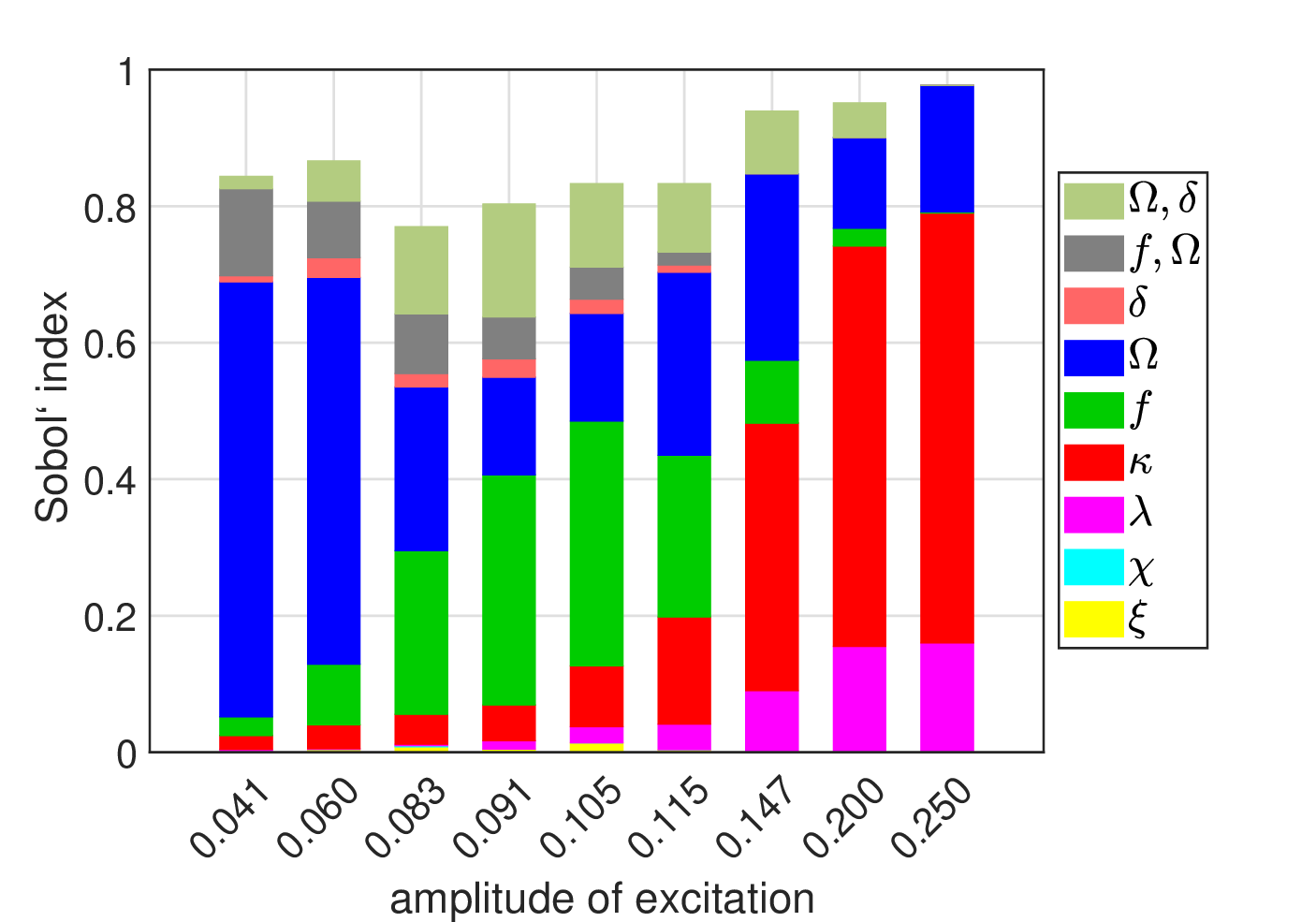}
    \caption{Sobol indices based on PCE method associated with the mean output power of the bistable energy harvester with asymmetric potential.}
    \label{fig:Sobol_asymmetric}
\end{figure}

First, the Sobol analysis is performed for the potential asymmetric configuration of the model, disregarding the bias angle, i.e., $\phi = 0$. Figure~\ref{fig:Sobol_asymmetric} presents the Sobol indices for the entire spectrum of amplitude. The quadratic nonlinearity term does not cause changes in the system's sensitivity. The influence of the excitation frequency at low amplitudes and the piezoelectric term at high amplitude influences are dominant. A noticeable portion of the variance comes from the second-order index related to the combined effect of $\delta$ and the excitation frequency. However, only the frequency has a significant first-order contribution.

\begin{figure}
    \centering
    \includegraphics[scale=0.36]{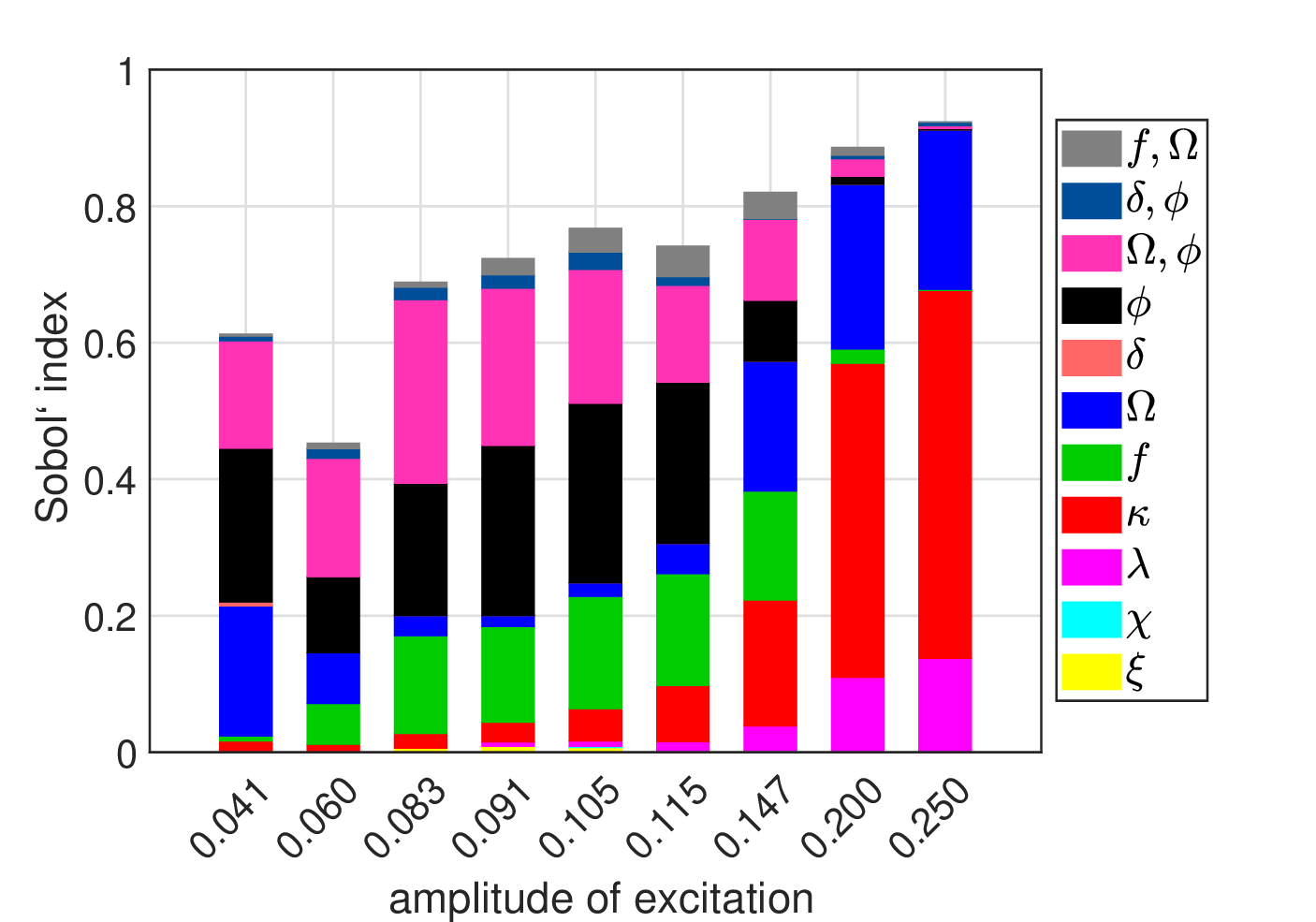}
    \caption{Sobol indices based on PCE method associated with the mean output power of the bistable energy harvester with asymmetric potential and bias angle.}
    \label{fig:Sobol_asymmetric_bias}
\end{figure}

Next, the bias angle is considered in the model. The bias angle $\phi$ is also described as a random variable with uniform distribution going from $-15^{\circ}$ to $15^{\circ}$. Positive angles show that the system undergoes a counterclockwise rotation, and the reverse occurs for negative angles. Figure~\ref{fig:Sobol_asymmetric_bias} shows the main Sobol indices. The bias angle significantly changes the system's sensitivity, especially for low amplitude cases, when the model exhibits unstable dynamic behavior. At low amplitudes, the model is more sensitive to variations in the bias angle $\phi$, excitation frequency $f$, and the joint effect (higher-order) from combinations of the parameters with the highest first-order indices. The frequency effect drops because of the bias angle higher effects, which become dominants. At high amplitude regions, the system increases its sensitivity to the piezoelectric coupling $\kappa$, but the asymmetry term $\delta$ keeps being irrelevant and can be its variability disregarded. Therefore, besides the excitation parameters and electrical piezoelectric properties, the bias angle also becomes an important parameter that can cause variability in the model.

\subsection{Asymmetric bistable energy harvester with nonlinear electromechanical coupling}

\begin{figure}
    \centering
    \includegraphics[scale=0.36]{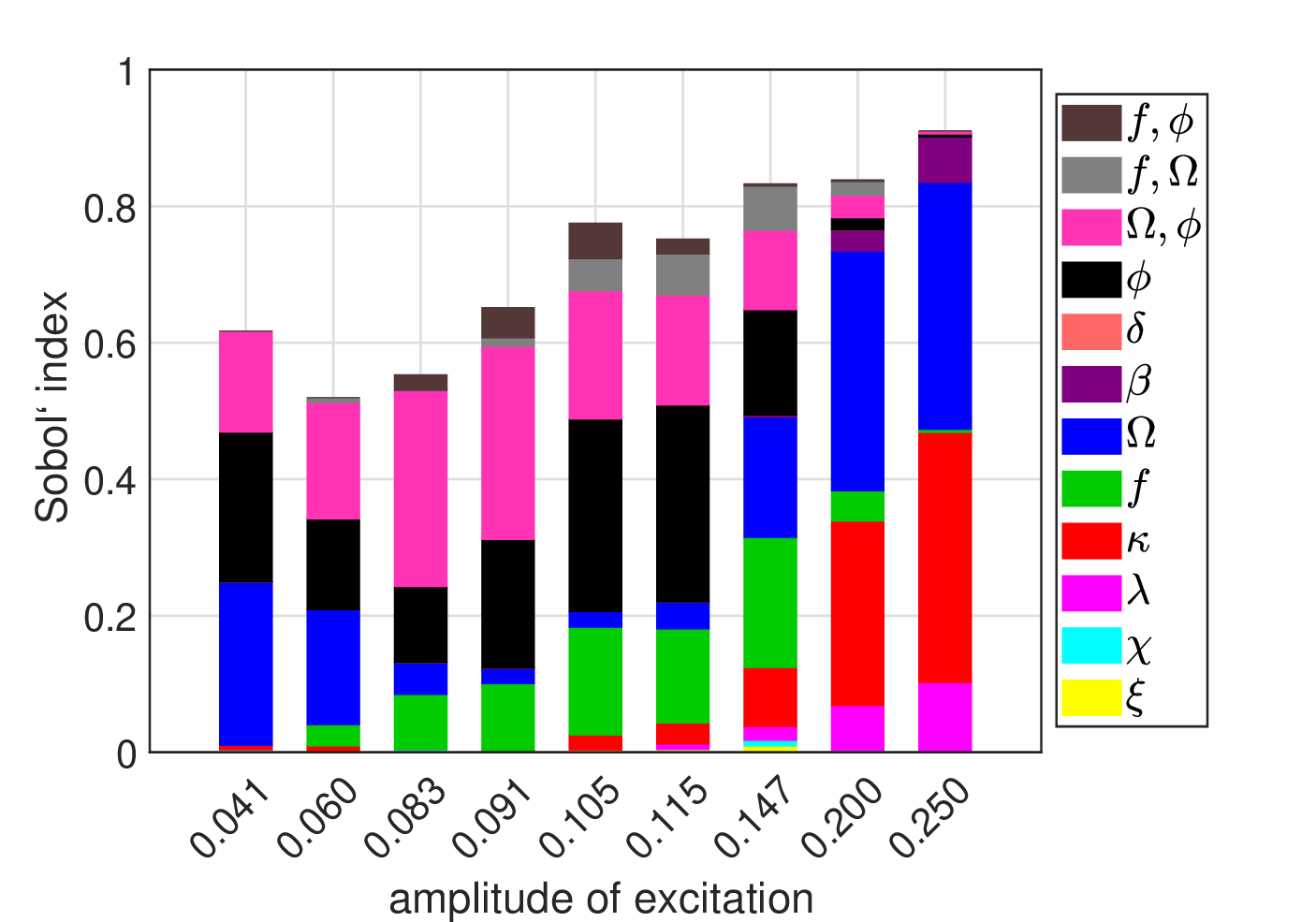}
    \caption{Sobol indices based on PCE method associated with the mean output power of the bistable energy harvester with asymmetric potential, bias angle, and nonlinear piezoelectric coupling.}
    \label{fig:Sobol_asymmetric_bias_Npzt}
\end{figure}

Finally, the proposed energy harvesting system considering asymmetry, bias angle, and nonlinear coupling is investigated. The range values of asymmetric coefficient and the bias angle are adopted as previously done. The nominal value of the nonlinear coupling is set equal to 1.0. Figure~\ref{fig:Sobol_asymmetric_bias_Npzt} displays the Sobol indices. The predominant effect from the bias angle, frequency and amplitude of excitation, and piezoelectric coupling in the electrical equation are observed. The effect of the nonlinear coupling does not demonstrate first or second-order influences for this scenario. Variability from the asymmetry coefficient, damping, and mechanical coupling terms do not affect the harvested power, so they might be deterministic terms. As well as in the previous cases, for low excitation amplitude, the system is sensitive to parameters that change its behavior, driven by the excitation frequency, amplitude, and bias angle. For the cases of the high amplitude of excitation, the first-order effect of the piezoelectric coupling is significant. Besides, the excitation frequency becomes the parameter with the greatest influence, progressively overcoming the influence of the piezoelectric coupling. This behavior occurs because of the nonlinear piezoelectric coupling, which demands more external energy for the system to achieve higher orbits, as already discussed. Thus, the frequency variations become very important in changing energy orbits, confirmed by the increase in its Sobol index.

Therefore, this complete nonlinear system is still sensitive to parameters that modify the system's orbit. The analysis is complex since the piezoelectric nonlinearity can quantitatively change the other parameters' sensitivity. However, its uncertain effects are only significant for high amounts.

\subsection{Summary of the findings}
\label{sec:summary}

Table~\ref{tab:tab2} summarizes the significant results obtained in this study. The results showed that the parameters with high first-order indices control the higher-order sensitivity terms. These results suggest that parameters with low sensitivity indices concerning the harvested power can be treated as fixed variables, reducing the dimension of the probabilistic model - these being the damping $\xi$, piezoelectric coupling in the mechanical equation $\chi$, and asymmetric coefficient $\delta$. In addition, identifying the parameters with low sensitivity indices allows for manufacturing constraints to be relaxed.

Parameters with medium and high sensitivity have a significant influence. Given the conditions, they should be treated as stochastic variables: the piezoelectric coupling in the electrical equation $\kappa$, external conditions $\Omega$ and $f$, bias angle $\phi$, nonlinear piezoelectric coupling $\beta$, and reciprocal time constant $\lambda$. Also, these parameters become relevant for optimization studies, such that minor changes in their values result in significant changes in the recovered power, which is essential for a proper preliminary optimization design. The innovative finding in this study is that the electrical propriety of piezoelectric coupling (piezoelectric coupling on the electrical equation) can be a starting point to an enhancement process.

\begin{table}[ht]
\centering

\caption{Summary for the system’s sensitivity to each parameter for the bistable energy harvesting models and their conditions to influence.}
\begin{tabular}{ccc}
\specialrule{.07em}{.05em}{.05em}\noalign{\smallskip}
\textbf{parameter} & \textbf{sensitivity} & \begin{tabular}[c]{@{}c@{}} \textbf{excitation} \\ \textbf{condition} \end{tabular} \\
\noalign{\smallskip}\specialrule{.07em}{.05em}{.05em}\noalign{\smallskip}

$\xi$     & none   & - \\
\noalign{\smallskip}\specialrule{.04em}{.05em}{.05em}\noalign{\smallskip}
$\chi$    & none   & - \\
\noalign{\smallskip}\specialrule{.04em}{.05em}{.05em}\noalign{\smallskip} 
$\lambda$ & low  & \begin{tabular}[c]{@{}c@{}} high amplitude \end{tabular}\\
\noalign{\smallskip}\specialrule{.04em}{.05em}{.05em}\noalign{\smallskip}
$\kappa$  & high & \begin{tabular}[c]{@{}c@{}} high amplitude \end{tabular}\\
\noalign{\smallskip}\specialrule{.04em}{.05em}{.05em}\noalign{\smallskip}
$f$       & high & \begin{tabular}[c]{@{}c@{}} low and middle \\ amplitude \end{tabular}\\
\noalign{\smallskip}\specialrule{.04em}{.05em}{.05em}\noalign{\smallskip}
$\Omega$  & high & \begin{tabular}[c]{@{}c@{}} all amplitudes \end{tabular}\\
\noalign{\smallskip}\specialrule{.04em}{.05em}{.05em}\noalign{\smallskip}
$\beta$   & low   & \begin{tabular}[c]{@{}c@{}} high $\beta$ values \end{tabular}\\
\noalign{\smallskip}\specialrule{.04em}{.05em}{.05em}\noalign{\smallskip}
$\delta$  & none   &  -\\
\noalign{\smallskip}\specialrule{.04em}{.05em}{.05em}\noalign{\smallskip}
$\phi$    & high  & \begin{tabular}[c]{@{}c@{}} low and middle \\ amplitude \end{tabular}\\
\noalign{\smallskip}\specialrule{.07em}{.05em}{.05em}
\end{tabular}
\label{tab:tab2} 
\end{table}

\begin{figure*}
    \includegraphics[width=0.98\textwidth]{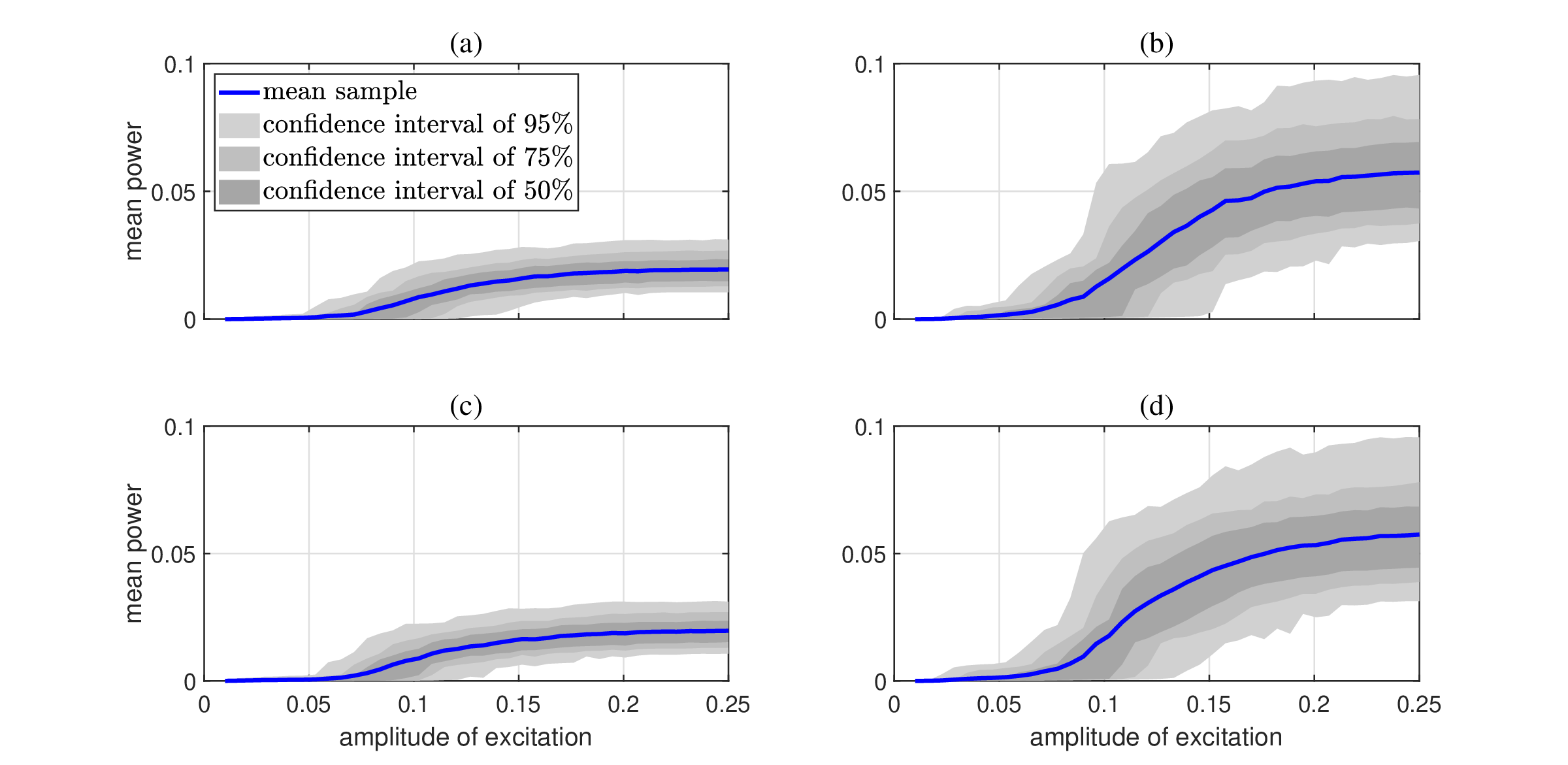}
    \caption{Uncertainty propagation of mean power in excitation amplitude spectrum: \textbf{(a)} Classical bistable energy harvester, \textbf{(b)} Classical bistable energy harvester with nonlinear coupling ($\beta=1$), \textbf{(c)} Asymmetric bistable energy harvester with linear coupling, \textbf{(d)} Asymmetric bistable energy harvester with nonlinear coupling ($\beta=1$).}
\label{fig:uq_plot}
\end{figure*}

Figure~\ref{fig:uq_plot} presents the uncertainty propagation of mean power in the excitation amplitude spectrum for the studied cases at different confidence levels. According to the table, it is noteworthy that uncertainties were only for the sensitivities parameters. Although remarkably nonlinear coupling generates a higher power, as already discussed, the uncertainty range is also higher than the case with nonlinear coupling, highlighting the need to verify the nonlinear term.

\section{Concluding remarks}
\label{sec:final remarks}

This work presented a global sensitivity analysis of a bistable energy harvesting system to identify the input parameters that lead to the most variability in the harvested power. Global sensitivity analysis based on Sobol indices was employed to calculate the parameters' individual and combined effect, exploring various dynamic behavior scenarios. We studied a classical bistable energy harvester, then added a nonlinear electromechanical coupling and, finally, a proposed model considering asymmetric coefficient at potential energy, slopping of the system's base, and the nonlinear coupling. All these considerations were done to study a sophisticated model, providing interesting insights for the proposal of a new bistable energy harvesting device.

The results here reaffirm the need to quantify and propagate uncertainties in bistable energy harvesting systems, presenting high sensitivity and a nontrivial interpretation by complex dynamic behavior. Furthermore, the framework proposes a powerful tool to develop a robust design, forecast, and optimization in bistable energy harvesting systems.

\section*{Acknowledgements}

The authors gratefully acknowledge, for the financial support given to this research, the following Brazilian agencies: Coordenação de Aperfeiçoamento de Pessoal de Nível Superior (CAPES) - Finance Code 001; São Paulo Research Foundation (FAPESP), grant number  19/19684-3; Brazilian National Council for Scientific and Technological Development (CNPq) grant number 306526/2019-0 ; Fundação Carlos Chagas Filho de Amparo à Pesquisa do Estado do Rio de Janeiro (FAPERJ) grants 210.167/2019, 211.037/2019 and 201.294/2021.

\section*{Code availability}

The simulations reported in this paper used the computational code {\bf STONEHENGE - Suite for Nonlinear Analysis of Energy Harvesting Systems} \cite{STONEHENGE_paper}. This code is available for free on GitHub \cite{STONEHENGE}.

\section*{Compliance with ethical standards}

\section*{Conflict of interest}
The authors declare they have no conflict of interest.

\bibliographystyle{spmpsci}      

\bibliography{references}

\begin{thebibliography}{10}
\providecommand{\url}[1]{{#1}}
\providecommand{\urlprefix}{URL }
\expandafter\ifx\csname urlstyle\endcsname\relax
  \providecommand{\doi}[1]{DOI~\discretionary{}{}{}#1}\else
  \providecommand{\doi}{DOI~\discretionary{}{}{}\begingroup
  \urlstyle{rm}\Url}\fi

\bibitem{27_Abbiati}
Abbiati, G., Marelli, S., Tsokanas, N., Sudret, B., B.~Stojadinovic, B.: A
  global sensitivity analysis framework for hybrid simulation.
\newblock Mechanical Systems and Signal Processing \textbf{146}, 106997 (2021).
\newblock \doi{10.1016/j.ymssp.2020.106997}

\bibitem{Alemazkoor}
Alemazkoor, N., Rachunok, B., Chavas, D., Staid, A., Louhghalam, A., Nateghi,
  R., Tootkaboni, M.: Hurricane-induced power outage risk under climate change
  is primarily driven by the uncertainty in projections of future hurricane
  frequency.
\newblock Scientific Reports \textbf{10}, 15270 (2020).
\newblock \doi{10.1038/s41598-020-72207-z}

\bibitem{26_Aloui}
Aloui, R., Larbi, W., Chouchane, M.: Global sensitivity analysis of
  piezoelectric energy harvesters.
\newblock Composite Structures \textbf{228}, 111317 (2019).
\newblock \doi{10.1016/j.compstruct.2019.111317}

\bibitem{Aloui_2020}
Aloui, R., Larbi, W., Chouchane, M.: Uncertainty quantification and global
  sensitivity analysis of piezoelectric energy harvesting using macro fiber
  composites.
\newblock Smart Materials and Structures \textbf{29}, 095014 (2020).
\newblock \doi{10.1088/1361-665X/ab9f12}

\bibitem{21_Arnold}
Arnold, D.: Review of microscale magnetic power generation.
\newblock IEEE Transactions on Magnetics \textbf{43}, 3940–3951 (2007).
\newblock \doi{10.1109/TMAG.2007.906150}

\bibitem{23_Cacuci}
Cacuci, D.: Sensitivity and uncertainty analysis: theory, vol.~1.
\newblock Boca Raton: Chapman Hall/CRC, New York (2003)

\bibitem{16_Cao}
Cao, J., Wang, W., Zhou, S., Inman, D., Lin, J.: Nonlinear time-varying
  potential bistable energy harvesting from human motion.
\newblock Applied Physics Letters \textbf{107}, 143904 (2015).
\newblock \doi{10.1063/1.4932947}

\bibitem{1_catacuzzeno}
Catacuzzeno, L., Orfei, F., {Di Michele}, A., Sforna, L., Franciolini, F.,
  Gammaitoni, L.: Energy harvesting from a bio cell.
\newblock Nano Energy \textbf{823-827}, 823 (2019).
\newblock \doi{10.1016/j.nanoen.2018.12.023}

\bibitem{7_Cottone}
Cottone, F., Vocca, H., Gammaitoni, L.: Nonlinear energy harvesting.
\newblock Physical Review Letters \textbf{102}, 080601 (2009).
\newblock \doi{10.1103/PhysRevLett.102.080601}

\bibitem{11_Crawley}
Crawley, E., Anderson, E.: Detailed models of piezoceramic actuation of beams.
\newblock Journal of Intelligent Material Systems and Structures \textbf{1},
  4--25 (1990).
\newblock \doi{10.1177/1045389X9000100102}

\bibitem{37_Crestaux}
Crestaux, T., {Le Maıˆtre}, O., Martinez, J.M.: Polynomial chaos expansion
  for sensitivity analysis.
\newblock Reliability Engineering \& System Safety \textbf{94}, 1161--1172
  (2009).
\newblock \doi{10.1016/j.ress.2008.10.008}

\bibitem{CunhaJR_CE}
{Cunha Jr}, A.: Enhancing the performance of a bistable energy harvesting
  device via the cross-entropy method.
\newblock Nonlinear Dynamics \textbf{103}, 137–155 (2021).
\newblock \doi{10.1007/s11071-020-06109-0}

\bibitem{CunhaJR_MC}
{Cunha Jr}, A., Nasser, R., Sampaio, R., Lopes, H., Breitman, K.: Uncertainty
  quantification through the monte carlo method in a cloud computing setting.
\newblock Computer Physics Communications \textbf{185}, 1355--1363 (2014).
\newblock \doi{10.1016/j.cpc.2014.01.006}

\bibitem{Daqaq}
Daqaq, M., Crespo, R., Ha, S.: On the efficacy of charging a battery using a
  chaotic energy harvester.
\newblock Nonlinear Dynamics \textbf{99}, 1525–1537 (2020).
\newblock \doi{10.1007/s11071-019-05372-0}

\bibitem{19_dAQAQ}
Daqaq, M., Masana, R., Erturk, A., Quinn, D.: On the role of nonlinearities in
  vibratory energy harvesting: A critical review and discussion.
\newblock Applied Mechanics Reviews \textbf{66}, 040801 (2014).
\newblock \doi{10.1115/1.4026278}

\bibitem{12_duToit}
duToit, N., Wardle, B.: Experimental verification of models for microfabricated
  piezoelectric vibration energy harvesters.
\newblock AIAAJ Journal \textbf{45} (2017).
\newblock \doi{10.2514/1.25047}

\bibitem{8_Erturk}
Erturk, A., Hoffmann, J., Inman, D.: A piezomagnetoelastic structure for
  broadband vibration energy harvesting.
\newblock Applied Physics Letters \textbf{94}, 254102 (2009).
\newblock \doi{10.1063/1.3159815}

\bibitem{22_Erturk}
Erturk, A., Inman, D.: An experimentally validated bimorph cantilever model for
  piezoelectric energy harvesting from base excitations.
\newblock Smart Materials and Structures \textbf{18}, 1–18 (2009).
\newblock \doi{10.1088/0964-1726/18/2/025009}

\bibitem{Erturk_valid}
Erturk, A., Inman, D.: Broadband piezoelectric power generation on high-energy
  orbits of the bistable duffing oscillator with electromechanical coupling.
\newblock Journal of Sound and Vibration \textbf{330}, 2339–2353 (2011).
\newblock \doi{10.1016/j.jsv.2010.11.018}

\bibitem{18_Franco}
Franco, V., Varoto, P.: Parameter uncertainties in the design and optimization
  of cantilever piezoelectric energy harvesters.
\newblock Mechanical Systems and Signal Processing \textbf{93}, 593–609
  (2017).
\newblock \doi{10.1016/j.ymssp.2017.02.030}

\bibitem{33_Ghanem}
Ghanem, R., Spanos, P.: Stochastic finite elements - a spectral approach.
\newblock Springer, Berlin (1991)

\bibitem{Halvorsen}
Halvorsen, E.: Fundamental issues in nonlinear wideband-vibration energy
  harvesting.
\newblock Physical Review E \textbf{87}, 042129 (2013).
\newblock \doi{10.1103/PhysRevE.87.042129}

\bibitem{He_asymmetric}
He, Q., M.F.Daqaq: Influence of potential function asymmetries on the
  performance of nonlinear energy harvesters under white noise.
\newblock Journal of Sound and Vibration \textbf{333}, 3479–3489 (2014).
\newblock \doi{10.1115/DETC2014-34397}

\bibitem{Hoeffding}
Hoeffding, W.: A class of statistics with asymptotically normal distribution.
\newblock Annals of Mathematical Statistics \textbf{19}, 293--325 (1948).
\newblock \doi{10.1214/aoms/1177730196}

\bibitem{30_Homma}
Homma, T., Saltelli, A.: Importance measures in the global sensitivity analysis
  of nonlinear models.
\newblock Reliability Engineering \& System Safety \textbf{52}, 1--17 (1996).
\newblock \doi{10.1016/0951-8320(96)00002-6}

\bibitem{17_Huang}
Huang, D., Zhou, S., Litak, G.: Nonlinear analysis of multistable energy
  harvesters for enhanced energy harvesting.
\newblock Communications in Nonlinear Science and Numerical Simulation
  \textbf{69}, 270–286 (2019)

\bibitem{2_inman}
Karami, M.A., Inman, D.J.: Powering pacemakers from heartbeat vibrations using
  linear and nonlinear energy harvesters.
\newblock Applied Physics Letters \textbf{100}, 042901 (2012).
\newblock \doi{10.1063/1.3679102}

\bibitem{Kroese_MC}
Kroese, D., Taimre, T., Botev, Z.I.: Handbook of Monte Carlo Methods, vol.~1.
\newblock John Wiley \& Sons, New Jersey (2011)

\bibitem{15_Leadenham}
Leadenham, S., Erturk, A.: Unified nonlinear electroelastic dynamics of a
  bimorph piezoelectric cantilever for energy harvesting, sensing, and
  actuation.
\newblock Nonlinear Dynamics \textbf{79}, 1727–1743 (2015).
\newblock \doi{10.1007/s11071-014-1770-x}

\bibitem{4_Lee}
Lee, Y., Qi, Y., Zhou, G., Lua, K.: Vortex-induced vibration wind energy
  harvesting by piezoelectric mems device information.
\newblock Scientific Reports \textbf{9}, 20404 (2019).
\newblock \doi{10.1038/s41598-019-56786-0}

\bibitem{9_Li}
Li, Y., Zhou, S., Litak, G.: Uncertainty analysis of bistable vibration energy
  harvesters based on the improved interval extension.
\newblock Journal of Vibration Engineering \& Technologies \textbf{8},
  297–306 (2020).
\newblock \doi{10.1007/s42417-019-00134-z}

\bibitem{Lopes}
Lopes, V., Peterson, J., {Cunha Jr}, A.: Nonlinear characterization of a
  bistable energy harvester dynamical system.
\newblock In: Topics in Nonlinear Mechanics and Physics, vol. 228, pp. 71--88.
  Springer, Singapore (2019).
\newblock \doi{10.1007/978-981-13-9463-8_3}

\bibitem{Lund}
Lund, A., Dyke, J., Song, W., Bilionis, I.: Global sensitivity analysis for the
  design of nonlinear identification experiments.
\newblock Nonlinear Dynamics \textbf{98}, 375–394 (2019).
\newblock \doi{10.1007/s11071-019-05199-9}

\bibitem{10_Mann}
Mann, B., Barton, D., Owens, B.: Uncertainty in performance for linear and
  nonlinear energy harvesting strategies.
\newblock Journal of Intelligent Material Systems and Structures \textbf{23},
  1451–1460 (2012).
\newblock \doi{10.1177/1045389X12439639}

\bibitem{20_Mitcheson}
Mitcheson, P., Miao, P., Stark, B., Yeatman, E., Holmes, A., Green, T.: Mems
  electrostatic micropower generator for low frequency operation.
\newblock Sensors and Actuators A: Physical \textbf{115}, 523–529 (2004).
\newblock \doi{10.1016/j.sna.2004.04.026}

\bibitem{Sudret2020PCA}
Nagel, J., Rieckermann, J., Sudrer, B.: Principal component analysis and sparse
  polynomial chaos expansions for global sensitivity analysis and model
  calibration: Application to urban drainage simulation.
\newblock Reliability Engineering \& System Safety \textbf{195}, 106737 (2020).
\newblock \doi{10.1016/j.ress.2019.106737}

\bibitem{STONEHENGE}
Norenberg, J., Peterson, J., Lopes, V., Luo, R., de~la Roca, L., Pereira, M.,
  Ribeiro, J., {Cunha Jr}, A.: {STONEHENGE} - {S}uite for {N}onlinear
  {A}nalysis of {E}nergy {H}arvesting {S}ystems (2021).
\newblock \urlprefix\url{https://americocunhajr.github.io/STONEHENGE}

\bibitem{STONEHENGE_paper}
Norenberg, J., Peterson, J., Lopes, V., Luo, R., de~la Roca, L., Pereira, M.,
  Ribeiro, J., {Cunha Jr}, A.: {STONEHENGE} — suite for nonlinear analysis of
  energy harvesting systems.
\newblock Software Impacts \textbf{10} (2021).
\newblock \doi{10.1016/j.simpa.2021.100161}

\bibitem{34_Oladyshkin}
Oladyshkin, S., Nowak, W.: Data-driven uncertainty quantification using the
  arbitrary polynomial chaos expansion.
\newblock Reliability Engineering \& System Safety \textbf{106}, 179--190
  (2012).
\newblock \doi{10.1016/j.ress.2012.05.002}

\bibitem{38_Palar}
Palar, P., Zuhal, L., Shimoyama, K., Tsuchiya, T.: Global sensitivity analysis
  via multi-fidelity polynomial chaos expansion.
\newblock Reliability Engineering \& System Safety \textbf{170}, 175--190
  (2018).
\newblock \doi{10.1016/j.ress.2017.10.013}

\bibitem{28_Ruiz}
Ruiz, R., Meruane, V.: Uncertainties propagation and global sensitivity
  analysis of the frequency response function of piezoelectric energy
  harvesters.
\newblock Smart Materials and Structures \textbf{26}, 065003 (2017).
\newblock \doi{10.1088/1361-665X/aa6cf3}

\bibitem{24_Saltelli}
Saltelli, A., Chan, K., Scott, E.: Sensitivity analysis.
\newblock Wiley, New York (2000)

\bibitem{35_SEPAHVAND}
Sepahvand, K., Marburg, S., Hardtke, H.: Uncertainty quantification in
  stochastic systems using polynomial chaos expansion.
\newblock Inter. J. App. Mech. \textbf{2}, 305--353 (2010).
\newblock \doi{10.1142/S1758825110000524}

\bibitem{29_Sobol}
Sobol, I.: Sensitivity estimates for nonlinear mathematical models.
\newblock Mathematical and Computer Modelling \textbf{1}, 407–414 (1993)

\bibitem{Soize2017}
Soize, C.: Uncertainty Quantification: An Accelerated Course with Advanced
  Applications in Computational Engineering, vol.~1.
\newblock Springer (2017)

\bibitem{14_Stanton}
Stanton, S., Erturk, A., Mann, B., Inman, D.: Nonlinear piezoelectricity in
  electroelastic energy harvesters: Modeling and experimental identification.
\newblock Journal of Applied Physics. \textbf{108}, 074903 (2010).
\newblock \doi{10.1063/1.3486519}

\bibitem{31_sudret}
Sudret, B.: Global sensitivity analysis using polynomial chaos expansions.
\newblock Reliability Engineering \& System Safety \textbf{93}, 964--979
  (2008).
\newblock \doi{10.1016/j.ress.2007.04.002}

\bibitem{13_triplett}
Triplett, A., Quinn, D.: The effect of nonlinear piezoelectric coupling on
  vibration-based energy harvesting.
\newblock Journal of Intelligent Material Systems and Structures \textbf{20},
  1959--1967 (2009).
\newblock \doi{10.1177/1045389X09343218}

\bibitem{Wang_enhancement}
Wang, W., Cao, J., Bowen, C., Zhang, Y., Lin, J.: Nonlinear dynamics and
  performance enhancement of asymmetric potential bistable energy harvesters.
\newblock Nonlinear Dynamics \textbf{94}, 1183–1194 (2018).
\newblock \doi{10.1007/s11071-018-4417-5}

\bibitem{3_wang}
Wang, Y., Yang, E., Chen, T., Wang, J., Hu, Z., Mi, J., Pan, X., Xu, M.: A
  novel humidity resisting and wind direction adapting flag-type triboelectric
  nanogenerator for wind energy harvesting and speed sensing.
\newblock Nano Energy \textbf{78}, 105279 (2020).
\newblock \doi{10.1016/j.nanoen.2020.105279}

\bibitem{6_Xin}
Xin, W., Zhang, Z., Huang, X., Hu, Y., Zhou, T., Zhu, C., Kong, X., Jiang, L.,
  Wen, L.: High-performance silk-based hybrid membranes employed for osmotic
  energy conversion.
\newblock Nature Communications \textbf{10}, 3876 (2019).
\newblock \doi{10.1038/s41467-019-11792-8}

\bibitem{36_Xiu}
Xiu, D.: Numerical Methods for Stochastic Computations: A spectral method
  approach.
\newblock Princeton University Press, Princeton (2010)

\bibitem{Yang_Fei}
Yang, K., Fei, F., An, H.: Investigation of coupled lever-bistable nonlinear
  energy harvesters for enhancement of inter-well dynamic response.
\newblock Nonlinear Dynamics \textbf{96}, 2369–2392 (2019).
\newblock \doi{10.1007/s11071-019-04929-3}

\bibitem{5_fang}
Yi, F., Wang, X., Niu, S., Li, S., Yin, Y., Dai, K., Zhang, G., Lin, L., Wen,
  Z., Guo, H., Wang, J., Yeh, M., Zi, Y., Liao, Q., You, Z., Zhang, Y., Wang,
  Z.: A highly shape-adaptive, stretchable design based on conductive liquid
  for energy harvesting and self-powered biomechanical monitoring.
\newblock Science Advances \textbf{2}, 1501624 (2016).
\newblock \doi{10.1126/sciadv.1501624}

\end{thebibliography}

\end{document}